\documentclass[twocolumn,prd,amsfonts,showpacs,epsfig,floats,nofootinbib,superscriptaddress,floatfix]{revtex4}
\usepackage{graphicx}
\usepackage{dcolumn}
\usepackage{bm}
\usepackage{longtable}
\usepackage{amsmath}

\begin{document}

\title{Cosmo-dynamics and dark energy with a quadratic EoS: \\ anisotropic models,  large-scale perturbations and cosmological singularities}


\author{ Kishore N. Ananda}
\affiliation{Institute of Cosmology and Gravitation, University of
Portsmouth, Mercantile House, Portsmouth PO1 2EG, Britain}
\author{Marco Bruni}
\affiliation{Institute of Cosmology and Gravitation, University of
Portsmouth, Mercantile House, Portsmouth PO1 2EG, Britain}
\affiliation{Dipartimento di Fisica, Universit\`a di Roma ``Tor
Vergata'', via della Ricerca Scientifica 1, 00133 Roma, Italy}
\date{\today}

\begin{abstract}
In standard general-relativistic cosmology, for fluids with a
linear equation of state (EoS) $P=w\rho$ or scalar fields, the
high isotropy of the universe requires special initial conditions:
singularities are velocity dominated and anisotropic in general.
In brane world effective 4-dimensional cosmological models an
effective term, quadratic in the energy density, appears in the
evolution equations, which has been shown to be responsible for
the suppression of anisotropy and inhomogeneities at the
singularity under reasonable assumptions. Thus in the brane world
isotropy is generically built in, and singularities are matter
dominated. There is no reason why the effective EoS of matter
should be linear at the highest energies,  and an effective
non-linear EoS may describe  dark energy  or  unified dark matter
(Paper I\,\cite{AB}).  In view of this, here  we investigate the
effects of a quadratic EoS in homogenous and inhomogeneous
anisotropic cosmological models in general relativity, in order to
understand if in this standard context the quadratic EoS can
isotropize the universe at early times. With respect to Paper
I\,\cite{AB}, here we use the simplified EoS $P=\alpha \rho +
\rho^2/\rho_c$, which still allows for an effective cosmological
constant and phantom behavior, and is general enough to analyze
the dynamics at high energies.  We first study homogenous and
anisotropic Bianchi I and V models, focusing on singularities.
Using dynamical systems methods, we find the  fixed points of the
system and study  their  stability. We find that models with
standard non-phantom behavior are  in general asymptotic in the
past to an isotropic fixed point IS, i.e. in these models even an
arbitrarily large anisotropy is suppressed in the past: the
singularity is matter dominated. Using covariant and gauge
invariant variables, we then study  linear anisotropic and
inhomogeneous perturbations about the homogenous and isotropic
spatially flat models with a quadratic EoS. We find that, in the
large scale limit, all perturbations decay asymptotically in the
past, indicating that the  isotropic fixed point IS is the general
asymptotic past attractor for non phantom inhomogeneous  models
with  a quadratic EoS.
\end{abstract}


\maketitle
\section{Introduction}\label{intro}

\indent Observations of large scale structure (LSS) \cite{LSS} and
cosmic microwave background radiation (CMBR)
\cite{CMBR,spergel,WMAP3} indicate that the universe is highly
homogenous and isotropic on large scales. Under standard
assumptions on the matter content, in the standard general
relativistic cosmological models isotropy is a special feature,
requiring a high degree of fine tuning (a special set of initial
conditions) in order to reproduce the observed universe.
Specifically, in the set of spatially homogeneous cosmological
models with  a fluid which satisfy the energy conditions $\rho
\geq |P|$ and $P\geq 0$, those which approach isotropy at late
times is of measure zero~\cite{CH}. In these models the anisotropy
dominates the dynamics in the neighborhood of the singularity,
while the effect of matter is negligible: thus the singularity can
be said to be velocity dominated \cite{ELS}. In general
anisotropic models do not isotropize sufficiently as they evolve
into the future (models do not evolve towards a
Friedmann-Robertson-Walker universe rapidly enough). This problem
is known as the isotropy problem and can be solved by inflation.
Many investigations on this issue show  that a large class of
models evolve towards a homogenous and isotropic model when under
the influence of inflationary matter. This means that  even models
with initially high levels of anisotropy can evolve into the
universe we observe today. A large body of work exists relating
specifically to the study of spatially homogenous and anisotropic
Bianchi cosmological models. In a classical paper, Wald showed
that initially expanding Bianchi models with a cosmological
constant asymptotically approach a de~Sitter state (hence a
spatially homogenous and isotropic model), except for the Bianchi
type IX mixmaster models which may re-collapse~\cite{RW}. In the
case of inhomogeneous models it is not so clear cut and in order
for inflation to begin, sufficiently homogenous initial conditions
are required (see e.g.~\cite{KT,BMP}). As far as singularities  in
general relativity are concerned, it is widely accepted
\cite{Garfinkle} that the  Belinskii, Khalatnikov and Lifshitz
(BKL) \cite{BKL,LL} picture is correct: for  fluids with linear
EoS $P= w \rho$ ($w<1$) or  scalar fields\footnote{Massless scalar
fields are equivalent to a stiff fluid $P=\rho$ ($w=1$) and are
special in this respect, see  \cite{Ashtekar} and references
therein.}, when the singularity of generic inhomogeneous
anisotropic cosmological models is approached, the dynamics is
asymptotically that of the mixmaster model, the singularity is
thus velocity dominated and the effect of matter is negligible
\cite{Ashtekar}.

More recently various theories of gravity that go beyond general
relativity have been considered. For instance, high energy
modifications of gravity could come from extra dimensions (as
required in string theory). In the brane
world~\cite{SMS,BMW,DL,RM} scenario the extra dimensions produce a
term quadratic in the energy density in the 4-dimensional
effective energy-momentum tensor, i.e.\ this quadratic term
appears in the 4-d  evolution and constraint equations on the
brane. Under the reasonable assumption of neglecting 5-d Weyl
tensor contributions in these equations, this quadratic term has
the very interesting effect of suppressing anisotropy at early
enough times, as the singularity is approached. In the case of a
Bianchi I brane-world cosmology containing a scalar field with a
large kinetic term the initial expansion is
quasi-isotropic~\cite{MSS}. In this scenario, the initial
anisotropy provides increased damping in the scalar field equation
of motion, thus resulting in an extended period of inflation. The
rapid decay of the anisotropy results in the class of initial
conditions allowed by observations to be greatly increased. Under
the same assumptions, Bianchi I and Bianchi V brane-world
cosmological models containing standard cosmological fluids with
linear equation of state (EoS) ($P/ \rho=w>0 $) also behave in a
similar manner~\cite{CS}. The same is also true for more general
homogeneous models \cite{coley1,coley2} and even some
inhomogeneous exact solutions \cite{coley3}. Finally, within the
limitations of a perturbative treatment, the
quadratic-term-dominated isotropic brane-world models have been
shown to be local past attractors in the larger state space of
inhomogeneous and anisotropic models~\cite{DGBC, GDCB}. More
precisely, again assuming that the 5-d Weyl tensor contribution to
the brane can be neglected, large-scale perturbations of the
isotropic models vanish  in the past, approaching the singularity.
Large scale perturbations $\lambda \gg H^{-1}$  are the only type
relevant to this kind of analysis for non-inflationary matter,
because for any given $\lambda$ there is always an early enough
time such that  $\lambda \gg H^{-1}$ is satisfied\footnote{We are
referring here to the well known fact that during a
non-inflationary (i.e.\ decelerated) phase $\ddot{a}<0$ a given
scale much larger than the Hubble horizon $\lambda\gg H^{-1}$ at a
given time will ``enter the horizon" ( $\lambda=H^{-1}$) at a
later time. In practice, neglecting small scales perturbations is
equivalent to neglect gradients and thus the Laplace operator term
in the evolution equations, which then become ordinary
differential equations. In the perturbative context, this is
analogous to the BKL approximation \cite{BKL,Ashtekar}.}. Thus in
the brane scenario the observed high isotropy of the universe is
built in, i.e.\ the natural outcome of {\it generic initial
conditions}, unlike in general relativity where general
cosmological models with a linear EoS are highly anisotropic in
the past, approaching the singularity {\it a-la} BKL, i.e. with
mixmaster dynamics \cite{BKL,LL,Ashtekar}.

It would therefore be interesting to see if the behavior of the
anisotropy at the singularity found in the brane scenario is
recreated in the general relativistic context if we consider a
quadratic term in the EoS. That is, do we have a scenario where,
in general relativity,  the generic asymptotic attractor in the
past is an isotropic model, thanks to a non-linear EoS. This is
the question we aim to investigate, for the specific case of the
EoS $P=\alpha \rho +  \rho^2/\rho_c$. This is a  simplified
version of that used in Paper I\,\cite{AB}, where the focus was on
using a quadratic EoS to represent a dark energy component or
unified dark matter. It is  general enough to analyze the  high
energy dynamics we are interested here, while still allowing for
an effective cosmological constant and phantom behavior.

The paper is organized as follows. In section II we outline the
setup of the models under investigation, defining three classes,
one of which represents a fluid with phantom behavior. Using
dynamical system methods \cite{WE,AP}, in section III we consider
the anisotropic Bianchi I and V class of homogenous models
containing a perfect fluid with quadratic EoS, showing how models
with  standard non-phantom matter are asymptotic in the past to an
isotropic fixed point IS. Using covariant and gauge invariant
variables \cite{EB,BDE,VEE}, in section IV we give the linearized
evolution and constraint equations for generic inhomogeneous and
anisotropic perturbations about a flat Friedmann model with a
quadratic EoS. We solve the equations in the large scale limit
$\lambda \gg H^{-1}$ and show that the perturbations vanish as
the singularity is approached, indicating that the isotropic model
IS is the generic past attractor. We then finish with some
concluding remarks and discussion in section V.

\section{The model}\label{section2}

\indent We begin with a summary of the most relevant points that can
be derived mostly from an analysis of energy conservation alone.
More details can be found in Section II in Paper I\,\cite{AB}.

The energy conservation equation for a
cosmological model containing a perfect fluid is:
\begin{eqnarray}
\label{conserv}
\dot{\rho}=-3H(\rho+P),
\end{eqnarray}

\noindent where $\rho$ is the energy density, $P$ the isotropic
pressure and $H$ is the Hubble expansion scalar. An average scale
factor $a$ is defined, at least locally, by $\dot{a}/{a}\equiv H$.
Through this relation solutions of Eq.\ (\ref{conserv}) can be found
in terms of $a$, but first an EoS must be specified. Usually this is
taken to be a linear function of the energy density of the form
$P=w\rho$, with $w=constant$. In this case Eq.\ (\ref{conserv})
gives:
\begin{eqnarray}
\rho &=& \rho_{o} \left(\frac{a}{a_o}\right)^{-3(w+1)}.
\end{eqnarray}

\noindent In the case of the homogenous and anisotropic Bianchi I
cosmological models, the amplitude of the shear scales as:
\begin{eqnarray}
\sigma^2 &=& \sigma_{o}^{2}\left(\frac{a}{a_o}\right)^{-6}.
\end{eqnarray}

\noindent This example shows how the shear can decay rapidly in an
expanding model, growing larger for smaller $a$. If the dynamics of
the model is such that $a\rightarrow 0$, both $\rho$ and $\sigma$
blow up, and the model has a  physical singularity. In the case $w<-1$, the
so called ``phantom" fluid \cite{Caldwell}, the energy density counter intuitively
decays (grows) in the past (future) and therefore any shear
component always comes to dominate at the initial ``Big Bang"
singularity. In the case when $-1\leq{\it w}<1$ the energy density
decays during expansion,  but at  a  slower rate than the shear, so
that once again the shear component dominates at the singularity. In
approaching these singularities matter with linear EoS $w<1$ is
un-influential, therefore these singularities may be termed velocity
dominated \cite{ELS}. It is only in the case when $w>1$
(``super-stiff" fluid) that the energy density is dominant in the
past and therefore the initial singularity is isotropic. This
indicates that as the initial singularity is approached in the
standard homogenous model with $w<1$, any residual anisotropy will
generally come to dominate near the singularity. The anisotropy is a
generic feature of these models and the initial singularity is in
general anisotropic and velocity dominated.

As discussed earlier, motivated by the results in the brane-world
models, we aim to investigate the effects of introducing a
non-linear EoS on the behavior of anisotropic models. The EoS we
propose is quadratic in the energy density (we neglect the constant
pressure term introduced in Paper I\,\cite{AB} as in general this only modifies
late time behavior):
\begin{eqnarray}
P=\alpha\rho + \frac{\epsilon\rho^2}{\rho_{c}}.
\end{eqnarray}

\noindent The discrete parameter $\epsilon$ denotes the sign of the
quadratic term, $\epsilon\in\{-1,1\}$, and $\rho_{c}>0$ sets the
energy scale above which this term becomes relevant. The energy
conservation equation (\ref{conserv}) can now be integrated to give
the scaling of the energy density as a function of the scale factor
(except for $\alpha=-1$).
\begin{eqnarray}\label{rhoA}
\rho &=& \frac{A(\alpha+1)\rho_{c}}{a^{3(\alpha+1)} - \epsilon A},\\
A &=& \frac{\rho_{o}  a_{o}^{3(\alpha+1)}}{(\alpha+1)\rho_{c}
+\epsilon \rho_{o}}.
\label{AA}
\end{eqnarray}

\noindent where $\rho_o$, $a_o$ represent the energy density and
scale factor at an arbitrary time $t_o$. This is valid for all
values of $\epsilon$, $\rho_{c}$ and $\alpha$ ($\alpha\neq-1$).
Defining
\begin{equation}
\rho_\Lambda \equiv -\epsilon (1+\alpha)\rho_c
\end{equation}

\noindent the fluid admits an effective positive cosmological
constant point, $\rho_\Lambda>0$, only if $\epsilon (1+\alpha) <0$.
As shown in Paper I\,\cite{AB}, in the case of $\epsilon=-1$ the fluid
generally behaves either in a phantom manner or is asymptotic to a
finite energy density  in the past, the effective cosmological
constant $\rho_\Lambda$, for all open and flat models. In this
scenario, the initial singularity is going to be dominated by
anisotropic expansion as seen in the standard linear EoS case, due
to the sub-dominance of the energy density at early times; therefore
we do not consider the $\epsilon=-1$ case further.

In the case of $\epsilon=+1$, the expression (\ref{rhoA}) for the
energy density $\rho(a)$ is conveniently rewritten in three
different ways, defining $a_{\star}= |A|^{1/3(\alpha+1)}$ and assuming $\rho>0$. \\

\noindent {\bf A:} $(1+\alpha)>0$, $\rho_\Lambda<0$,
\begin{equation}
\rho=\frac{(1+\alpha)\rho_c}{\left(\frac{a}{a_\star}\right)^{3(1+\alpha)}
-1}. \label{rho1}
\end{equation}

\noindent In this case $a_\star<a<\infty$, with $\infty>\rho >0$.
The past singularity is similar to certain future singularities
found in dark energy models~\cite{NOT, barrow} and  is referred to
as Type III singularities:
\begin{itemize}
\item For $t \to t_{\star}$, $a \to a_{\star}$,
$\rho \to \infty$ and $|P| \to \infty$
\end{itemize}

\noindent Where $t_{\star}$ and $a_{\star}$ are constants with
$a_{\star}\neq 0$ (we reach $a_*$ at a finite $t_*$). The main
difference is that in our case the singularity occurs in the past.
At late times the fluid behaves in a similar manner to a fluid
with a linear EoS and $w=\alpha$.\\

\noindent {\bf B:} $(1+\alpha)<0$, $0<\rho_\Lambda<\rho$,
\begin{equation}
\rho=\frac{\rho_\Lambda}{1-\left(\frac{a}{a_\star}\right)^{3(1+\alpha)}}.
\label{rho2}
\end{equation}

\noindent In this case $a_\star<a<\infty$, with $\infty>\rho
>\rho_\Lambda$. As in case {\bf A}, the fluid approaches a Type
III singularity in the past, but now approaches an effective
cosmological constant at late times.\\

\noindent {\bf C:} $(1+\alpha)<0$, $0<\rho<\rho_\Lambda$,
\begin{equation}
\rho=\frac{\rho_\Lambda}{1+\left(\frac{a}{a_\star}\right)^{3(1+\alpha)}}.
\label{rho3}
\end{equation}

\noindent In this case $0<a<\infty$, with $0<\rho <\rho_\Lambda$.
The fluid behaves in a phantom manner but tends to an effective
cosmological constant at late times.\\

\noindent In the cases {\bf A} and {\bf B}, as we approach the past
singularity the energy density approaches infinity, in fact the
singularity occurs at a finite scale factor, that is for the energy
density one has $\rho \to \infty$ as $a \to a_{\star}$. The
cosmological model emerges at a finite scale factor $a_{\star}$,
with infinite energy density and infinite isotropic pressure $P$.
The model emerges from a curvature singularity, infinitely
decelerates and asymptotically approaches a standard flat Friedmann
model with a linear EoS (case {\bf A}) or a de Sitter model (case
{\bf B}). In the case {\bf C}, the fluid behaves in a phantom manner
and approaches $\rho_{\Lambda}$ at late times. In this case the
energy density is sub-dominant at early times and the anisotropy
will dominate as we will show in the proceeding sections. In cases
{\bf A} and {\bf B}, we have the possibility that the energy density
could be the dominant component in the past, forcing the singularity
to be isotropic. This would result in isotropic expansion at early
times as seen in the brane-world model, with the fluid also
approaching a non super-stiff EoS at late times. We will now proceed
with the analysis of anisotropic cosmological models containing a
fluid with a quadratic EoS.

\section{Dynamics of Bianchi models}\label{section3}

We now study the dynamics of homogeneous and anisotropic Bianchi
cosmological models. In particular we investigate the Bianchi I and
V classes of models that contain the flat and open Friedmann models.
The matter content will be modeled by a non-tilted (that is the
fluid flow remains orthogonal to the group orbits) perfect fluid
with a quadratic EoS.

We make use of the orthonormal frame formalism which results in a
system of first-order evolution equations. We have adopted a
notation consistent with that used in~\cite{WE} (Greek indices run
from $\alpha=1,\ldots,3$ and Latin indices run from $i=0,\ldots,3$).
The orthonormal frame formalism is a 1+3 decomposition of the
Einstein field equations into evolution and constraint equations.
The decomposition is carried out relative to the time-like vector
field $ {\bf{e}}_{0} $ of a given orthonormal frame
$\{{\textbf{e}}_{i}\}$. In the case of models which admit an
isometry group, $ {\bf {e}}_{0} $ is chosen to be the normal (${\bf
u}$) to the space-like group orbits. The Bianchi models are such
cosmological models, they admit a three dimensional group of
isometries which act simply and transitively on the space-like
hyper-surfaces (surfaces of homogeneity). The orthonormal frame
$\{{\bf u},{\bf e}_{\alpha}\}$, is such that:
\begin{eqnarray}
\textbf{u} \cdot \textbf{u} = -1,\qquad \textbf{u} \cdot
{\textbf{e}}_{\alpha} = 0,\qquad {\textbf{e}}_{\alpha} \cdot
{\textbf{e}}_{\beta}= \delta_{\alpha\beta}.
\end{eqnarray}

\noindent The dynamics can then be described in terms of the
spatial commutation functions, the kinematical quantities and the
matter variables. The spatial commutation functions,
$\gamma^{\alpha}{}_{{\beta}{\mu}}$, are defined by the commutation
relations between the spatial basis vectors such that:
\begin{equation}
[{\textbf{e}}_{\alpha},{\textbf{e}}_{\beta}]=\gamma^{\mu}{}_{
{\alpha}{\beta}}{\textbf{e}}_{\mu}.
\end{equation}

\noindent The spatial commutation functions can be decomposed into
a $2$-index symmetric object and a $1$-index object which
determine the connections on the space-like 3-surfaces.
\begin{equation}
\gamma^{\alpha}{}_{{\beta}{\mu}} =
\varepsilon_{{\beta}{\mu}{\nu}}n^{{\alpha}{\nu}} +
a_{\beta}\delta_{\mu}{}^{\alpha} -
a_{\mu}\delta_{\beta}{}^{\alpha}.
\end{equation}

\noindent The kinematical quantities include the Hubble scalar $H$,
the shear tensor $\sigma_{{\alpha}{\beta}}$ and the angular velocity
$\Omega_{\alpha}$, relative to a Fermi-propagated spatial frame, of
the spatial frame $\{{\textbf{e}}_{\alpha}\}$ along the time-like
vector ${\bf u}$. The relevant matter variables are the energy
density $\rho$ and the isotropic pressure $P$, while the energy flux
and the anisotropic pressure are set to zero as we are considering a
perfect fluid. The decomposed Einstein field equations  for a
general Bianchi model containing a perfect fluid then take the form:
\begin{eqnarray}
\dot{H} &=& -H^2 - \frac{2}{3}\sigma^2 - \frac{1}{6}(\rho + 3P),\label{EFE1}\\
{\dot{\sigma}}_{{\alpha}{\beta}} &=& -3H
{\sigma}_{{\alpha}{\beta}} - ^{(3)}{S_{{\alpha}{\beta}}} \nonumber\\
&& + 2\Omega_{\nu} (\varepsilon^{{\mu}{\nu}}{}_{{\alpha}}
\sigma_{{\beta}{\mu}} + \varepsilon^{{\mu}{\nu}}{}_{{\beta}}
\sigma_{{\alpha}{\mu}}),\label{EFE2}\\
\rho &=& 3H^2 - {\sigma}^2 +  \frac{1}{2} {^{(3)}{R}},\label{EFE3}\\
0 &=& 3{\sigma}_{\alpha}{}^{\beta}a_{\beta} -
\varepsilon_{\alpha}{}^{{\beta}{\mu}} {\sigma}_{\beta}{}^{\nu}
n_{{\nu}{\mu}}.\label{EFE4}
\end{eqnarray}

\noindent The trace-free spatial Ricci tensor
(${}^{3}{S_{{\alpha}{\beta}}} =
{}^{3}{R_{{\alpha}{\beta}}}-\frac{1}{3}~
{}^{3}{R}\delta_{\alpha\beta}$) and the spatial curvature scalar
(${}^{3}{R} = {}^{3}{R_{\alpha}^{\alpha}}$) can be given in terms
of the spatial commutation functions:
\begin{eqnarray}
^{3}{S_{{\alpha}{\beta}}} &=& b_{{\alpha}{\beta}} -
\frac{1}{3}(b_{\mu}{}^{\mu})\delta_{{\alpha}{\beta}} \nonumber\\
&& - 2 a_{\nu} (\varepsilon^{{\nu}{\mu}}{}_{{\alpha}}
n_{{\beta}{\mu}} + \varepsilon^{{\nu}{\mu}}{}_{{\beta}} n_{{\alpha}{\mu}}),\label{curv1}\\
^{3}{R} &=& -\frac{1}{2} b_{\mu}{}^{\mu} - 6
a_{\mu}a^{\mu}.\label{curv2}
\end{eqnarray}

\noindent The quantity $b_{{\alpha}{\beta}}$, can be expressed in
terms of the spatial commutation functions and has the form:
\begin{equation}
b_{{\alpha}{\beta}} = 2 n_{\alpha}{}^{\mu} n_{{\mu}{\beta}}
-n_{\mu}{}^{\mu} n_{{\alpha}{\beta}}.\label{curv3}
\end{equation}

\noindent The Jacobi identity for vector fields when applied to
the spatial frame vectors ${\textbf{e}}_{\alpha}$, in conjunction
with the Einstein field equations give a set of evolution and constraints equations:
\begin{eqnarray}
{\dot{n}}_{{\alpha}{\beta}} &=& -H{n}_{{\alpha}{\beta}} +
2(\sigma_{\alpha}{}^{\mu} n_{{\beta}{\mu}} +
\sigma_{\beta}{}^{\mu} n_{{\alpha}{\mu}}) \nonumber\\
&& + 2\Omega_{\nu} (\varepsilon^{{\mu}{\nu}}{}_{{\alpha}}
\sigma_{{\beta}{\mu}} +
\varepsilon^{{\mu}{\nu}}{}_{{\beta}} \sigma_{{\alpha}{\mu}}),\label{evo1}\\
{\dot{a}}_{\alpha} &=& -H a_{\alpha} - {\sigma}_{\alpha}{}^{\beta}
a_{\beta} +
\varepsilon_{\alpha}{}^{{\mu}{\nu}}a_{\mu}\Omega_{\nu},\label{evo2}\\
0 &=& n_{\alpha}{}^{\beta} a_{\beta}.\label{evo3}
\end{eqnarray}

\noindent The evolution equation for the energy density results from
the contracted Bianchi identities, or more directly from the
conservation equations, and has the form (\ref{conserv}). The latter
with the above system of equations reduce to the evolution and
constraint equations for given non-tilted perfect fluid Bianchi
model by making further restrictions on the spatial commutation
functions ($n_{\alpha\beta}$,$a_{\alpha}$) and kinematical
quantities ($\sigma_{\alpha\beta}$,$\Omega_{\alpha}$). The Bianchi I
and V models are considered separately in the following
sub-sections.

\subsection{Bianchi I cosmologies}

The Bianchi I models are a subclass of the Bianchi Class A models
($a_{\alpha}=0$). These models are homogeneous and anisotropic
cosmological models containing the flat Friedmann (F) model. In
the case of Bianchi I, the group invariant orthonormal frame
$\{{\bf u  },{\bf e }_{\alpha}\}$ can be further specialized such
that:
\begin{eqnarray}
a_{\alpha}=n_{\alpha\beta} &=&0,\\
\gamma^{\mu}{}_{\alpha\beta} &=&0.
\end{eqnarray}

\noindent This in turn means that the three curvature of the spatial
hyper-surfaces (\ref{curv1}, \ref{curv2}) is zero:
\begin{eqnarray}
^{(3)}{R}_{\alpha\beta} &=& 0.
\end{eqnarray}

\noindent The Einstein field equations (\ref{EFE2}, \ref{EFE4}) and
the Jacobi identity (\ref{evo1}) imply:
\begin{eqnarray}
\sigma_{{\alpha}{\beta}} =
diag(\sigma_{11},\sigma_{22},\sigma_{33}),\qquad
 \Omega_{\alpha}= 0.
\end{eqnarray}

\noindent The analysis of the system of equations can be simplified
by introducing the following set of normalized dimensionless
variables:
\begin{eqnarray}
x = \frac{\rho}{\rho_{c}}~,\qquad
y=\frac{H}{\sqrt{\rho_{c}}}~,\nonumber
\end{eqnarray}
\begin{eqnarray}
z = \frac{\sigma}{\sqrt{\rho_{c}}}~,\qquad
\eta=\sqrt{\rho_{c}}t~,
\end{eqnarray}

\noindent where $\sigma\equiv \sqrt{  \sigma_{\alpha\beta}
\sigma^{\alpha\beta}  /2  }$. In addition, it is customary to
introduce
\begin{equation}
\label{Sigma}
\Sigma=\frac{\sigma}{H}=\frac{z}{y}
\end{equation}

\noindent as a dynamically dimensionless measure of anisotropy: we
can refer to a cosmological model as isotropic in some limit if, in
that limit, $\Sigma \rightarrow 0$. In particular, if there is a
singularity for some $a_s$ (possibly zero), we refer to the
singularity as isotropic if $\Sigma \rightarrow 0$ in the limit
$a\rightarrow a_s$.

\noindent The above normalization to $\rho_c $ is useful because the
energy scale  $\rho_c$ itself disappears from the dynamical system:
in terms of the dimensionless variables above, the complete system
of equations for Bianchi I models is explicitly made scale
invariant, and is:
\begin{eqnarray}
y^{2} &=& \frac{x}{3} + \frac{z^2}{3},  \label{genfried}  \\
x'&=& -3 y \left( (\alpha +1)x +  x^2 \right) ,\\
y' &=& -y^{2} - \frac{1}{6} \left( (3\alpha +1)x + 3 x^{2} \right) ,\\
z'&=& -3 y z.
\end{eqnarray}

\noindent The primes denote differentiation with respect to $\eta$,
the normalized time variable. The variable $x$ is the normalized
energy density, $y$ is the normalized Hubble scalar and $z$ is the
normalized shear. The effective cosmological constant point is now
given by $x_\Lambda =-(1+\alpha)$. We only consider the region of
the state space for which the energy density remains positive
($x\geq0$) and models which are expanding ($y\geq0$). The evolution
equation for the shear can be integrated to give:
\begin{eqnarray}\label{shearevo}
z = z_{o} \left( \frac{a}{a_{o}} \right)^{-3}~.
\end{eqnarray}

\noindent The shear tends to grow in the past and rapidly decays at
late times. Using Eqs. (\ref{rho1}), (\ref{rho2}), (\ref{rho3}) and
(\ref{shearevo}) we can see how the shear scales with respect to the
energy density (by substituting for the scale factor in each case)
for each of the three sub-cases of the fluid.\\

\noindent Case {\bf A:} $(1+\alpha)>0$, $\rho_\Lambda<0$,
\begin{equation}
z=z_{o} \left( \frac{x}{|\alpha+1|+x}
\right)^{\frac{1}{|\alpha+1|}}.
\end{equation}

\noindent In this case, at early times the energy density diverges,
$x\to\infty$ and the shear goes to a constant value, $z \to z_{o}$.
The initial singularity is dominated by the energy density and is
forced to be isotropic, as $\Sigma \to 0$.
At late times, $x \to 0$ and $z \to 0$.\\

\noindent Case {\bf B:} $(1+\alpha)<0$, $0<\rho_\Lambda<\rho$,
\begin{equation}
z=z_{o} \left( \frac{x-x_{\Lambda}}{x}
\right)^{\frac{1}{|\alpha+1|}}.
\end{equation}

\noindent In this case, at early times the energy density diverges,
$x\to\infty$; the shear goes to a constant, $z \to z_{o}$. The
initial singularity is again dominated by the energy density and is
forced to be isotropic, with $\Sigma \rightarrow 0$. At late times
the energy density approaches a constant
value, $x \to x_{\Lambda}$ and $z \to 0$.\\

\noindent Case {\bf C:} $(1+\alpha)<0$, $0<\rho<\rho_\Lambda$,
\begin{equation}
z=z_{o} \left( \frac{x_{\Lambda}-x}{x}
\right)^{\frac{1}{|\alpha+1|}}.
\end{equation}

\noindent In the final case, the fluid behaves in a phantom manner.
At early times  $x\to 0$ and  $z \to \infty$. The initial
singularity is now dominated by the shear and is anisotropic. At
late times the energy density approaches a
constant value, $x \to x_{\Lambda}$ and $z \to 0$.\\

In cases {\bf A} and {\bf B} we expect the generic past attractor to
be an isotropic singularity, while at late times we expect either a
Minkowski or de Sitter model to be the attractor. In case {\bf C},
the fluid behaves in a phantom manner and we expect models to be
asymptotic to a anisotropic singularity in the past and a de Sitter
model in the future. We now carry out a dynamical systems analysis
of the above system. The system can be reduced from a 3D system to a
2D system by substituting for a variable using the generalized
Friedmann equation (\ref{genfried}). We choose to remove the
normalized Hubble  scalar $y$, assuming expansion, $y>0$. The
resulting reduced planar dynamical system is then:
\begin{eqnarray}
x'&=& -\sqrt{3(x+z^2)} \left( (\alpha +1)x +  x^2 \right) , \label{xprimeI}\\
z'&=& -\sqrt{3(x+z^2)} z. \label{zprimeI}
\end{eqnarray}

The above system of equations is of the form $u'_{i}=f_{i}(u_{j})$.
Since this system is autonomous, trajectories in state space connect
the fixed/equilibrium points of the system ($u_{j,o}$), which
satisfy the system of equations $f_{i}(u_{j,0})=0$. The stability
nature of the fixed points can be found by carrying out a linear
stability analysis, see Section II.B in Paper I\,\cite{AB} for a summary, and
e.g.\ \cite{WE, AP} for more  details. The fixed points of the
reduced planar system (\ref{xprimeI})-(\ref{zprimeI}), the
corresponding eigenvalues and there existence conditions (the energy
density remains positive ($x\geq0$) and all three variables are real
($x,~y,~z\in\mathbb{R}$)) are given in Table~\ref{Tab1}.

\begin{center}
\begin{table}[h!]\caption{\label{Tab1}Location, eigenvalues and
existence conditions ($x\geq0$ and $x,~y,~z\in\mathbb{R}$) for the
fixed points of the reduced Bianchi I planar system
(\ref{xprimeI})-(\ref{zprimeI}). To simplify we use
$\tau=-(\alpha+1)$.}
\begin{tabular*}{0.47\textwidth}{@{\extracolsep{\fill}}ccccc}
\hline \hline
\\
Name & $x$,~$z$  & Eigenvalues & Existence \\
\hline
\\
M & ($0$,~$0$) &  $0$,~$0$ & $\alpha\in\mathbb{R}$ \\
\\
dS & ($\tau$,~$0$) & $-\tau\sqrt{3\tau}$,~$-\sqrt{3\tau}$ & $\alpha<-1$  \\
\\
\hline \hline
\end{tabular*}
\end{table}
\end{center}

\begin{figure*}
\begin{center}
\includegraphics[height=8.5cm,width=8.5cm, angle=270]{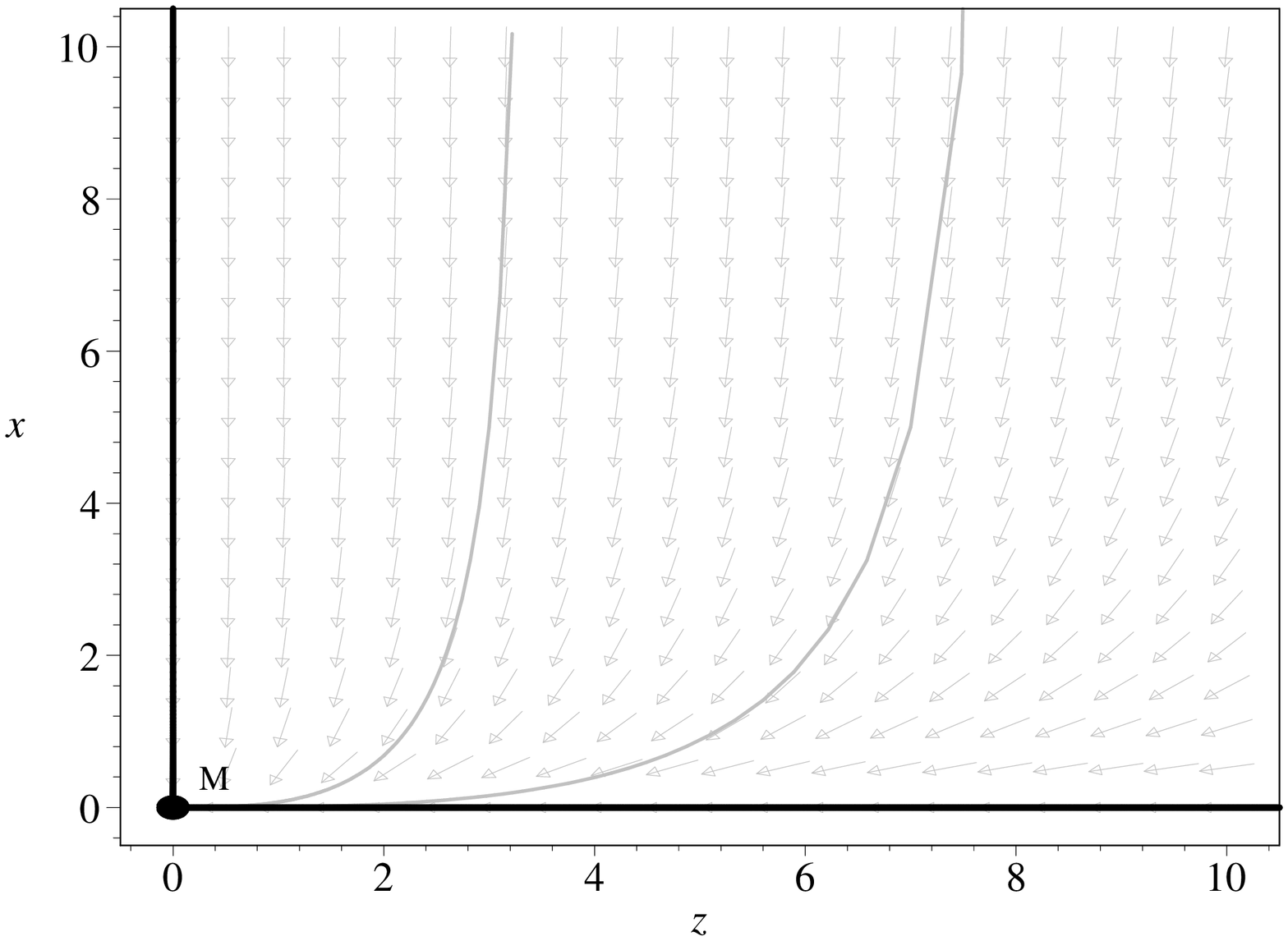}
\includegraphics[height=8.5cm,width=8.5cm, angle=270]{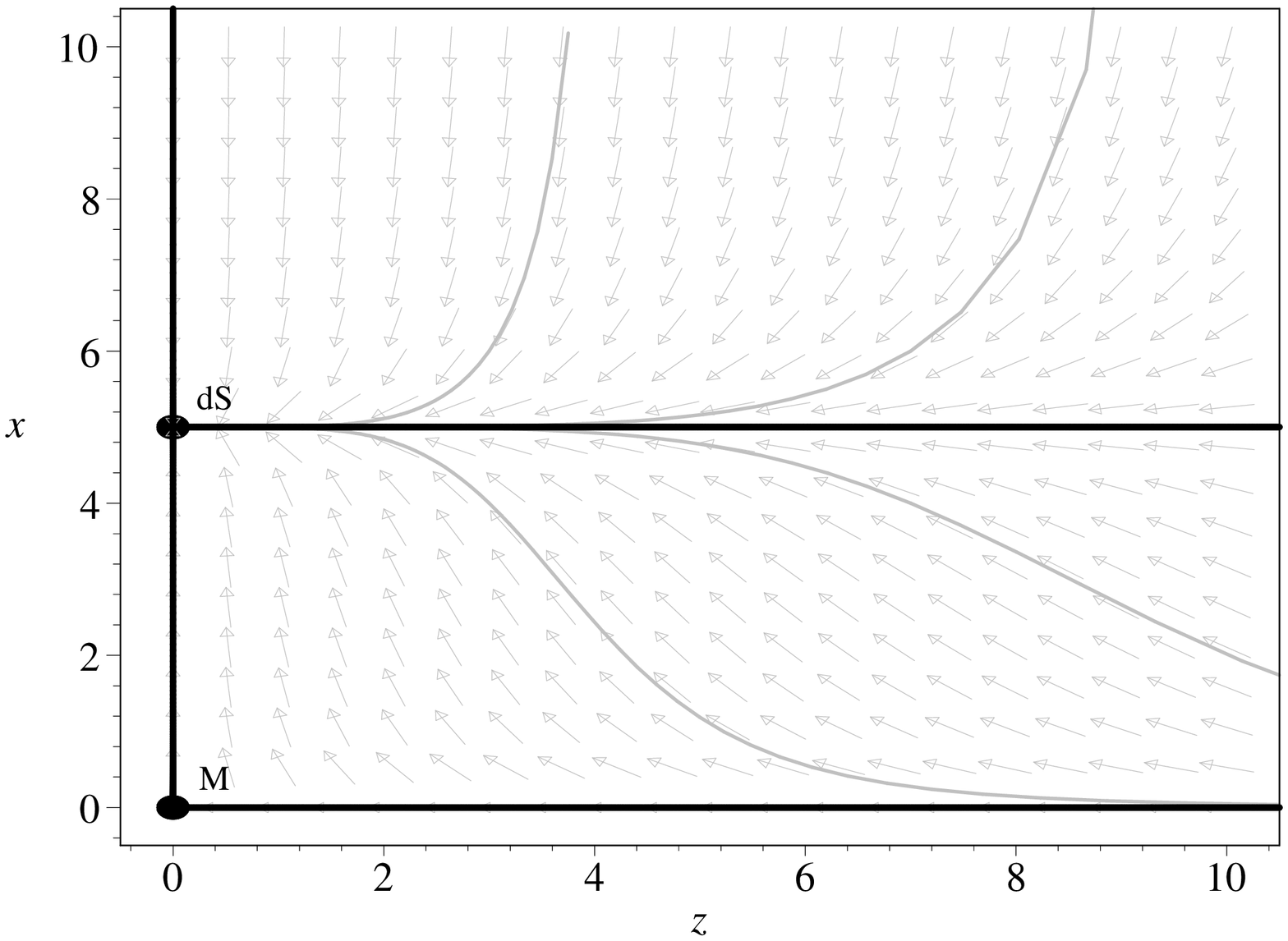}
\caption{The state space for the Bianchi I reduced planar system
(\ref{xprimeI})-(\ref{zprimeI}), showing fixed points at finite
values of the energy density and shear. (a) Left panel: the state
space for $\alpha>-1$; the only fixed point present represents a
Minkowski model. (b) Right panel: the state space for $\alpha<-1$;
in this case there is a new fixed point dS which represents a flat
expanding de Sitter model with $\rho_\Lambda=-(1+\alpha)\rho_c$. The
separatrix represents models with $\rho=\rho_\Lambda$. Models below
this line have phantom behavior.  }\label{fig1}
\end{center}
\end{figure*}

\noindent The first point represents the vacuum static Minkowski
model. Eigenvalues of the linearized system are zero, hence the
linearization theorem  \cite{AP} doesn't apply at this point and we
resort to numerical analysis to determine the stability character.
In the case $\alpha>-1$, this is the only finite valued fixed point
present. This is  clearly seen  in Fig.~\ref{fig1}(a), corresponding
to case {\bf A}, where the black lines represent separatrices, the
dots represent fixed points and the thin grey lines are example
trajectories. The Minkowski point M is clearly the future attractor
for all trajectories in the region. The behavior in the past is not
clear from this diagram (but trajectories do appear to evolve to a
constant $z$). The horizontal black line represents the Bianchi I
Kasner vacuum models, with  line element:
\begin{eqnarray}\label{Kas}
ds^2 = -dt^2 + t^{2p_{1}}dx_{1}^{2}+ t^{2p_{2}}dx_{2}^{2}+
t^{2p_{3}}dx_{3}^{2}.
\end{eqnarray}

\noindent The parameters $p_{\alpha}$ satisfy the following
constraints:
\begin{eqnarray}
\sum_{n=1}^{3} p_{n} =1,\qquad \sum_{n=1}^{3} p_{n}^2 =1,
\end{eqnarray}

\noindent and the average scale factor for expanding models of this
type is given by:
\begin{eqnarray}
a=a_{o}(\eta-\eta_{o})^{1/3}.
\end{eqnarray}

\noindent Although these models are asymptotic to Minkowski at late
times, as seen in Fig.\  \ref{fig1}(a), they are always anisotropic,
in the sense that $\Sigma=\sqrt{3}$ at all times. The Kasner
singularity is a paradigmatic, characteristic of many models; the
more general Bianchi IX singularity is approximated by a series of
jumps from a Kasner phase to another, see e.g.\ \cite{LL,WE}.
The vertical black line represents the expanding flat
Friedmann model with a quadratic EoS. In the case of a spatially
flat model as in this case, the Friedmann equation can be
integrated to give the dependence of the scale factor on time:
\begin{eqnarray}
\eta &=& \frac{6}{[3(\alpha+1)]^{3/2}} \times
\left\{\sqrt{(a/a_{\star})^{3(\alpha+1)} -1}\right.\nonumber\\
\nonumber\\
&&\left. -\arctan\left(\sqrt{{(a/a_{\star})^{3(\alpha+1)}
-1}}\right) \right\}
\end{eqnarray}

\noindent The scale factor starts  at  finite size $a_{\star}$ (we
have fixed constants so that this corresponds to $\eta_{\star}=0$),
expands and asymptotically approaches the behavior of a model with
the standard linear EoS,
\begin{eqnarray}
a\approx a_{\star}(\eta-\eta_{o})^{2/3(\alpha+1)}.
\end{eqnarray}

\noindent For $\alpha<-1$, Fig.\ \ref{fig1}(b), there is a  second
fixed point dS representing the generalized flat expanding de Sitter
model. This point  has attractor stability character, i.e.\ in this
case the generalized de Sitter model is the generic future
attractor.  The new horizontal black line in Fig.~\ref{fig1}(b) is a
separatrix representing models with $x_{\Lambda}=-(\alpha+1)$; it is
the dividing line between the phantom ($x<x_{\Lambda}$) and standard
behavior ($x>x_{\Lambda}$). The trajectories below this line
represent phantom models (corresponding to case {\bf C}), that is
the energy density increases as the model expands. The models in
this region appear to be dominated by shear in the past ($x \to 0$,
$ z \to \infty$ as $\eta \to -\infty$) and asymptotically approach a
de Sitter model in the future ($x\to x_{\Lambda})$, $z \to 0$ as
$\eta \to \infty$). The scale factor for an expanding flat de Sitter
model has the form:
\begin{eqnarray}\label{ds}
a=a_{o}e^{(\eta-\eta_{o})}       .
\end{eqnarray}

\noindent The trajectories above the separatrix (corresponding to
case {\bf B}) also asymptotically approach the generalized de Sitter
model in the future.

\begin{figure*}[t]
\begin{center}
\includegraphics[height=8.5cm,width=8.5cm, angle=270]{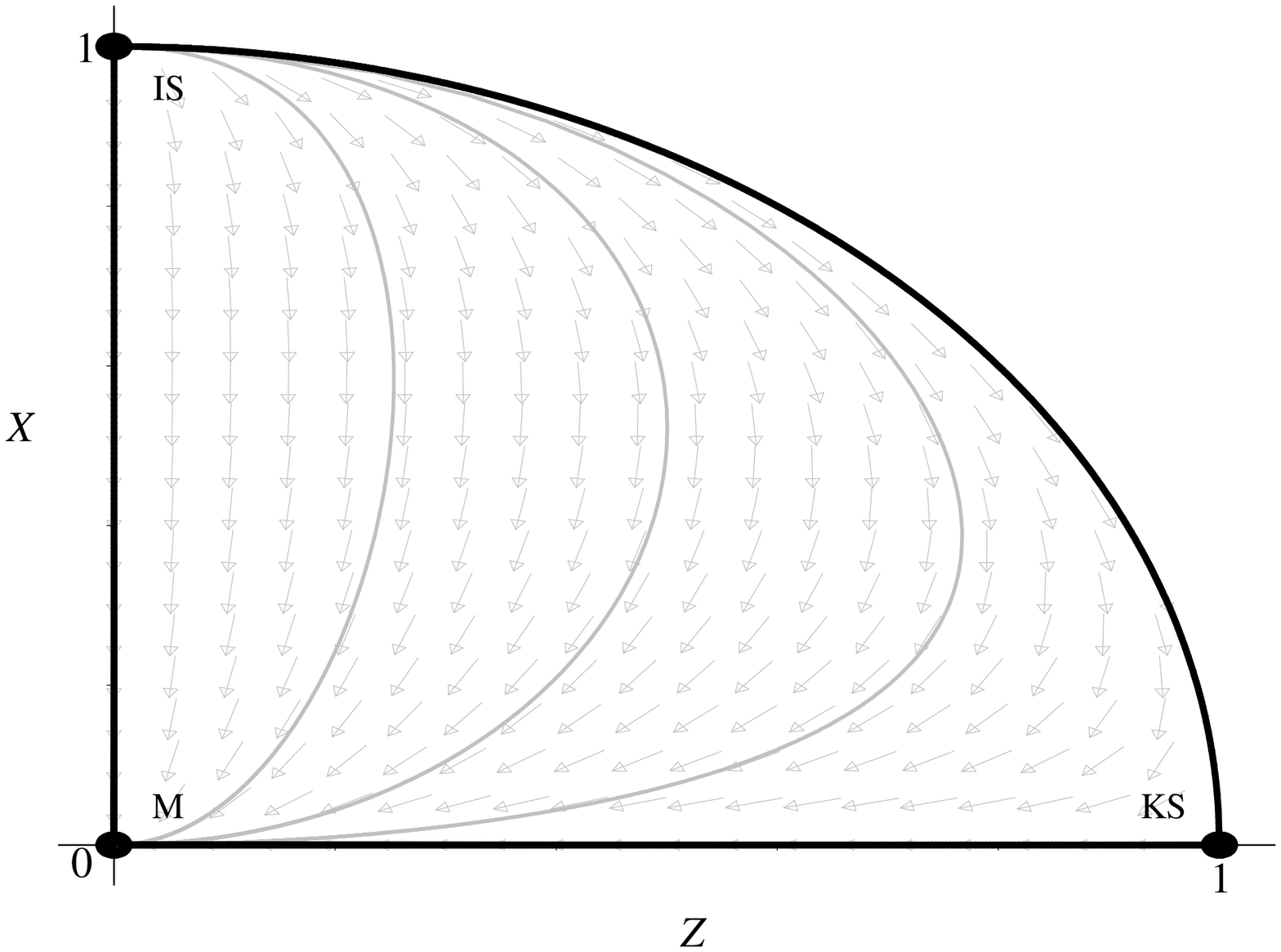}
\includegraphics[height=8.5cm,width=8.5cm, angle=270]{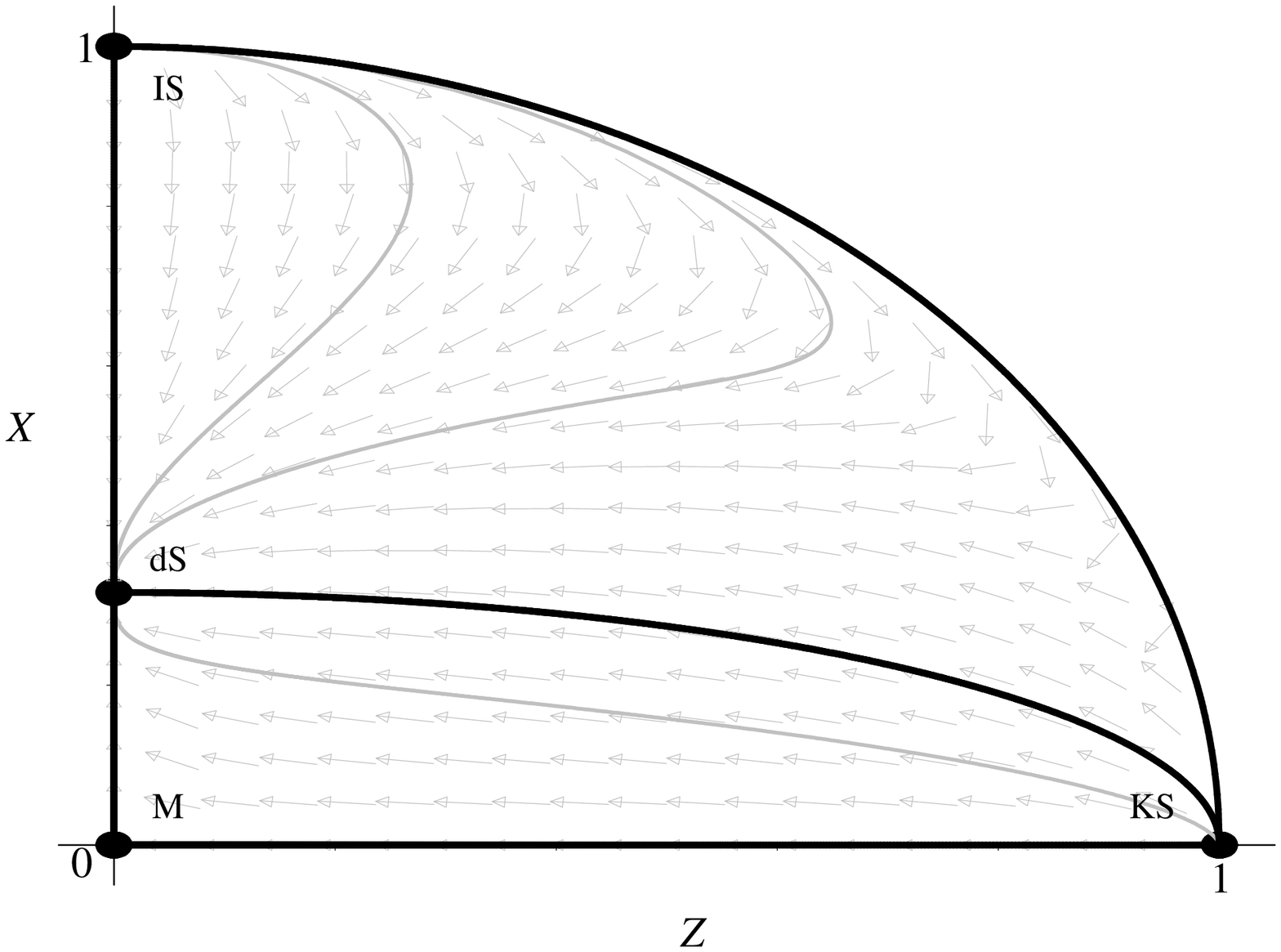}
\caption{The compactified state space ($X$,~$Z$) for the Bianchi I
reduced system, including fixed points at infinity. These represent
the isotropic singularity (IS) and the Kasner anisotropic
singularity (KS). (a) Left panel: the state space for $\alpha>-1$,
corresponding to case {\bf A}, Fig.\ \ref{fig1}(a). (b) Right panel:
the state space for $\alpha<-1$. The separatrix corresponds to that
of Fig.\ 1(b), with $\rho=\rho_{\Lambda}$. Above this line we have
models of type {\bf B}. Models below this line are phantom, of type
{\bf C}, and Kasner-like at the singularity KS. Clearly the
non-phantom models of type {\bf A} and {\bf B} are asymptotic to the
IS fixed point in the past.}\label{fig2}
\end{center}
\end{figure*}

The behavior in the past is not clear from the state space in
Fig.~\ref{fig1}. In order to analyze the dynamics at infinity, we
introduce compactified Poincar\'{e}  variables ${X}_{i}$; these can
be expressed in terms of the standard variables $x_{i}$ as:
\begin{eqnarray}
{X}_{i} &=& \frac{x_{i}}{\sqrt{ 1 + {\sum_{i=1}^{n} x_{i}^{2}}} }.
\end{eqnarray}

\noindent The relationship can be inverted to give:
\begin{eqnarray}
x_{i} &=& \frac{{X}_{i}}{\sqrt{ 1 - {\sum_{i=1}^{n}
{X}_{i}^{2}}}}.
\end{eqnarray}

\noindent The new variables allow one to analyze the dynamics at
infinity more easily, as $x_{i} \to \infty \Rightarrow {X}_{i} \to
1$. The evolution equations for the new variables are then given by:
\begin{eqnarray}
{X}_{i}' &=& \sqrt{ 1 - {\sum_{j=1}^{n} {X}_{j}^{2}} } \left(
x_{i}' - {X}_{i}{\sum_{k=1}^{n}~{X}_{k}~x_{k}' } \right),
\end{eqnarray}

\noindent where the $x_{i}'$ represent the evolution equations for
the original variables. We now study the compactified and reduced
system ($X$,~$Z$); the fixed points of the system are given in
Table~\ref{Tab2}.

The first two fixed points (M and dS) represent the Minkowski and
generalized de Sitter points as in the non-compactified case and
retain the same stability character. The third point (IS) represents
the isotropic singularity, which is matter dominated ($X \to 1$, or
$x \to \infty$). This fixed point always has repeller stability and
is the generic past attractor. The final fixed point (KS) represents
the anisotropic Kasner singularity ($Z \to 1$, or $z \to \infty$).
This fixed point has saddle stability. The full state space for the
system can now be reconstructed, and is depicted in Fig.\
\ref{fig2}, where as before the black lines represent separatrices,
the dots represent fixed points and the thin grey lines are example
trajectories.

\begin{center}
\begin{table}[h!]\caption{\label{Tab2} Location and existence conditions
($X\geq0$ and $X,~Z\in\mathbb{R}$) for the compactified planar
Bianchi I system. To simplify we use $\tau=-(\alpha+1)$. The * is
shown when the point does not exist.}
\begin{tabular*}{0.47\textwidth}{@{\extracolsep{\fill}}cccc}
\hline \hline
\\
Name & $X$,~$Z$  & Existence & Stability ($\alpha<-1,~\alpha>-1$) \\
\hline
\\
M & ($0$,~$0$) & $\alpha\in\mathbb{R}$ & Saddle,~Attractor \\
\\
dS & $\left( \frac{\tau}{\sqrt{1+\tau^2}},~0\right)$ & $\alpha<-1$ & Attractor,~*   \\
\\
IS & ($1$,~$0$) & $\alpha\in\mathbb{R}$ & Repeller,~Repeller   \\
\\
KS & ($0$,~$1$) & $\alpha\in\mathbb{R}$  & Saddle,~Saddle \\
\\
\hline \hline
\end{tabular*}
\end{table}
\end{center}

\noindent In the case $\alpha>-1$, Fig.~\ref{fig2}(a) (this
corresponds to case {\bf A}, i.e. this is the compactified version
of Fig.~\ref{fig1}(a)), there are three fixed points, M, IS and KS.
The horizontal black line $X=0$ represents the anisotropic Kasner
model. The vertical black line $Z=0$ represents the flat Friedmann
model containing a fluid with a quadratic EoS. The Minkowski point M
is the future attractor for all trajectories in the region. The
isotropic singularity IS is the generic past attractor and the
anisotropic Kasner singularity KS has saddle stability character.
The set of initial conditions for which anisotropy dominates in the
past are special, corresponding to vacuum models $X=0$, as opposed
to being generic as in the case of the {\it linear} EoS.

In the case $\alpha<-1$, Fig.\ \ref{fig2}(b), there are four fixed
points, M, dS, IS and KS.  The state space is split into two regions
by the separatrix joining the points dS and KS and given by
$X=X_{\Lambda}$, where:
\begin{eqnarray}
X_{\Lambda} &=& \tau\sqrt{\frac{1-Z^2}{1+\tau^2}}.
\end{eqnarray}

\noindent and $\tau=-(\alpha+1)$. In the region above the separatrix
($X>X_{\Lambda}$, corresponding to case {\bf B}) the fluid behaves
in a standard manner. The IS fixed point has repeller stability
character and all trajectories are asymptotic to it in the past. The
dS fixed point has attractor stability and is the generic future
attractor. The KS fixed point has saddle stability in this region.
In the region below the separatrix ($X<X_{\Lambda}$, corresponding
to case {\bf C}) the fluid behaves in a phantom manner. The KS fixed
point has repeller stability character and the dS fixed point is an
attractor. The M fixed point has saddle stability character. In
general, phantom trajectories in this region are asymptotic to the
Kasner singularity in the past and the generalized de Sitter model
in the future.

\subsection{Bianchi V cosmologies}

The models of type Bianchi V are a subclass of the Bianchi Class B
models ($a_{\alpha} \neq 0$). These are homogeneous and anisotropic
cosmological models containing both the flat Friedmann (F) model and
the open Friedmann models (the Milne model for $\rho=0$). In the
case of Bianchi V models the group invariant orthonormal frame
$\{{\bf u},{\bf e }_{\alpha}\}$ can again be specialized further:
\begin{eqnarray}
a_{2}=a_{3}=n_{\alpha\beta} &=&0,\\
a_{1} &\neq&0,\\
\gamma^{\mu}{}_{\alpha\beta} &=&0.
\end{eqnarray}

\noindent This in turn means that the spatial curvature scalar is:
\begin{eqnarray}
^{3}{R} &=& -6a_{1}^{2}.
\end{eqnarray}

\noindent Where the evolution equation for $a_{1}$ is:
\begin{eqnarray}
\dot{a}_{1} &=& -H a_{1}.
\end{eqnarray}

\noindent The Einstein field equation (\ref{EFE4}) implies:
\begin{eqnarray}
\sigma_{{\alpha}1} = 0.
\end{eqnarray}

\noindent We will continue to use the normalized variables
introduced in the previous subsection, $x$, $y$, $z$ and $\eta$.
Additionally we introduce the new normalized curvature variable
$c$:
\begin{eqnarray}
c &=& \frac{a_{1}}{\sqrt{|\rho_{c}|}}.
\end{eqnarray}

\noindent The generalized Friedmann equation is now:
\begin{eqnarray}
y^{2} &=& \frac{x}{3} + \frac{z^2}{3} + c^2.
\end{eqnarray}

\noindent As with the previous case, this constraint equation can be
used to reduce the dimension of the dynamical system: again we chose
to remove $y$. The resulting set of evolution equation is then:
\begin{eqnarray}
x'&=& - \sqrt{3x + 3z^2 + 9c^2} \left( (\alpha +1)x +  x^2 \right) ,\\
\nonumber\\
z'&=& -z \sqrt{3x + 3z^2 + 9c^2} ,\\
\nonumber\\
c' &=& -c \sqrt{\frac{x}{3} + \frac{z^2}{3} + c^2}.
\end{eqnarray}

\noindent The fixed points of the new reduced system, the
corresponding eigenvalues and their existence conditions (the energy
density remains positive, $x\geq0$, and all four variables are real,
$x,~y,~z,~c\in\mathbb{R}$) are given in Table~\ref{Tab3}. The fixed
points are identical to those found in the Bianchi I case. As with
the Bianchi I case, in order to analyze the dynamics at infinity, we
introduce the Poincar\'{e} variables ${X}_{i}$. We then study the
compactified and reduced system ($X$,~$Z$,~$C$), with the fixed
points given in Table~\ref{Tab4}.

\begin{center}
\begin{table}[h!]\caption{\label{Tab3}Location, eigenvalues and
existence conditions ($x\geq0$ and $x,~y,~z,~c\in\mathbb{R}$) for
the finite value fixed points of the reduced Bianchi V system. To
simplify we use $\tau=-(\alpha+1)$.}
\begin{tabular*}{0.47\textwidth}{@{\extracolsep{\fill}}ccccc}
\hline \hline
\\
Name & $x$,~$z$,~$c$  & Eigenvalues & Existence \\
\hline
\\
M & ($0$,~$0$,~$0$) &  $0$,~$0$,~$0$ & $\alpha\in\mathbb{R}$ \\
\\
dS & ($\tau$,~$0$,~$0$) & $-\tau\sqrt{3\tau}$,~$-\sqrt{3\tau}$,~$-\sqrt{\tau/3}$ & $\alpha<-1$  \\
\\
\hline \hline
\end{tabular*}
\end{table}
\end{center}

\begin{center}
\begin{table}[h!]\caption{\label{Tab4}Location and existence conditions
($X\geq0$ and $X,~Z,~C\in\mathbb{R}$) for the compactified and
reduced Bianchi V system. To simplify we use $\tau=-(\alpha+1)$. The
* is shown when the point does not exist.}
\begin{tabular*}{0.47\textwidth}{@{\extracolsep{\fill}}cccc}
\hline \hline
\\
Name & $X$,~$Z$,~$C$  & Existence & Stability ($\alpha<-1,~\alpha>-1$) \\
\hline
\\
M & ($0$,~$0$,~$0$) & $\alpha\in\mathbb{R}$ & Saddle,~Attractor \\
\\
dS & $\left( \frac{\tau}{\sqrt{1+\tau^2}},~0,~0\right)$ & $\alpha<-1$ & Attractor,~*   \\
\\
CS & ($0$,~$0$,~$1$) & $\alpha\in\mathbb{R}$ & Saddle,~Saddle   \\
\\
IS & ($1$,~$0$,~$0$) & $\alpha\in\mathbb{R}$ & Repeller,~Repeller   \\
\\
KS & ($0$,~$1$,~$0$) & $\alpha\in\mathbb{R}$  & Saddle,~Saddle \\
\\
\hline \hline
\end{tabular*}
\end{table}
\end{center}

\begin{figure*}
\begin{center}
\includegraphics[height=8.5cm,width=8.5cm, angle=0]{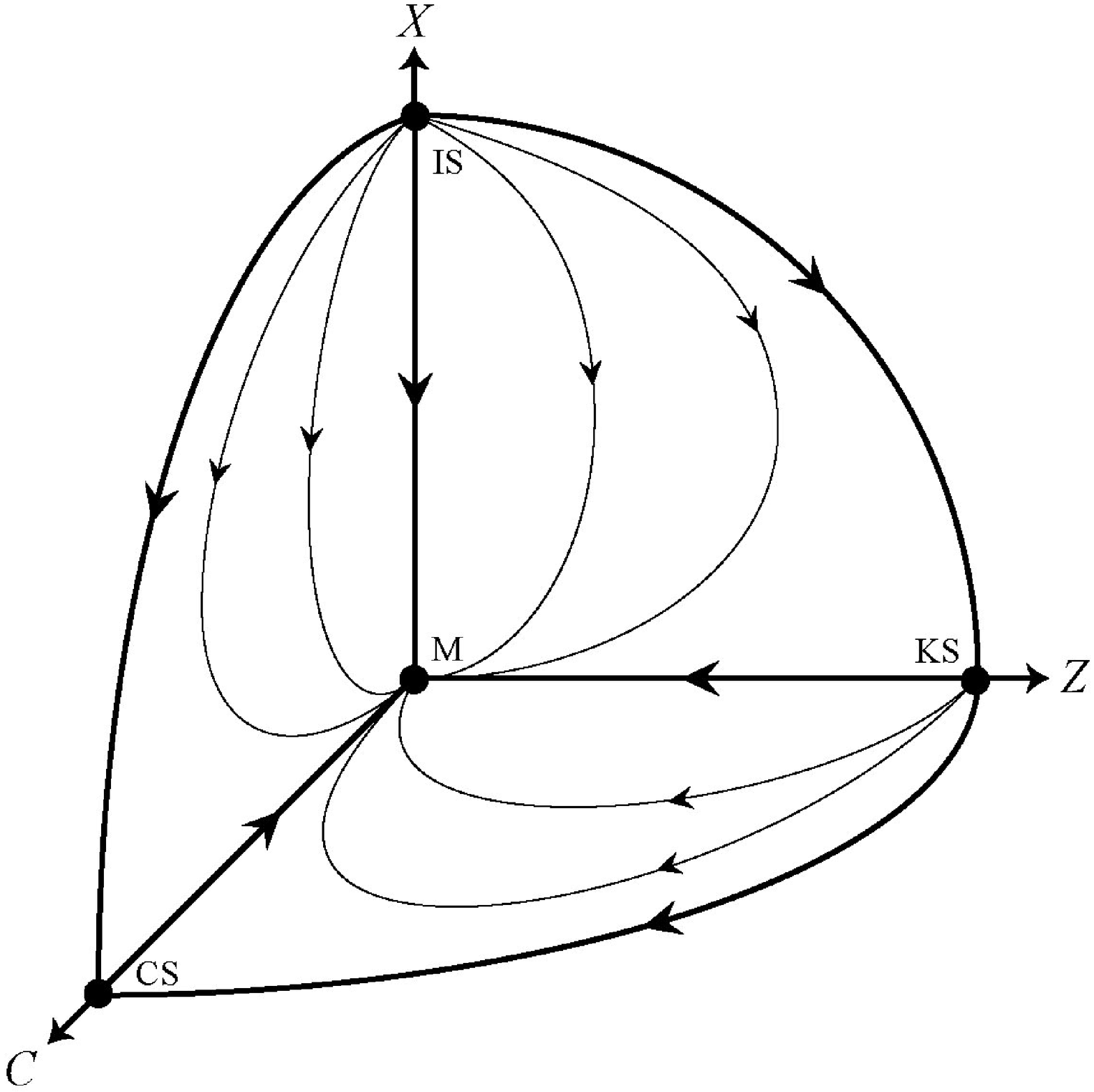}
\includegraphics[height=8.5cm,width=8.5cm, angle=0]{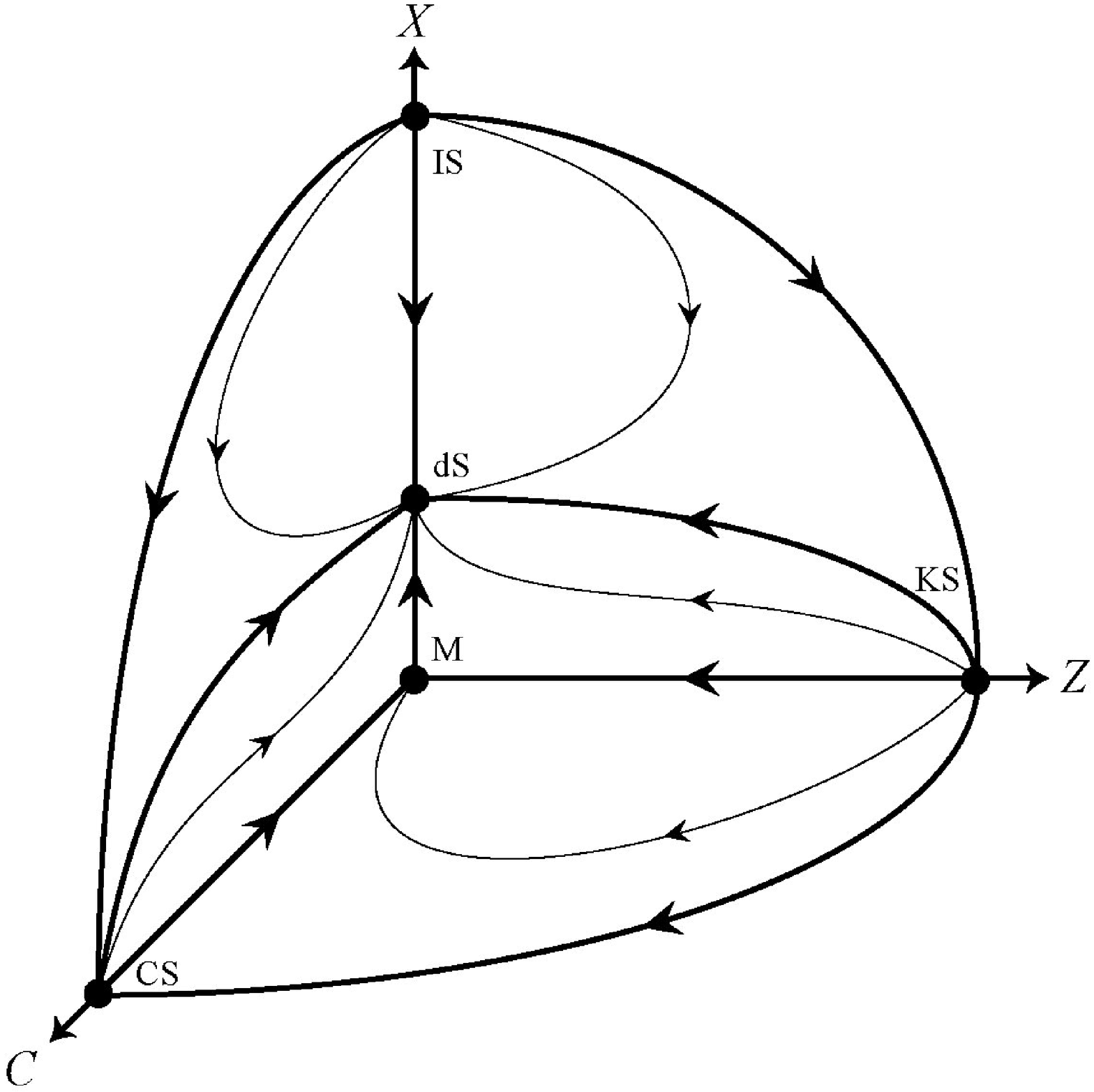}
\caption{The compactified state space ($X$,~$Z$,~$C$) for the
Bianchi V reduced system, including fixed points at infinity. These
represent the isotropic singularity (IS), the Kasner anisotropic
singularity (KS) and the curvature singularity (CS). (a) Left panel:
The state space when $\alpha>-1$. (b) Right panel: the state space
when $\alpha<-1$. It is clear that standard (non phantom)
trajectories are asymptotic to the IS fixed point in the past.
}\label{fig3}
\end{center}
\end{figure*}

\noindent The first two fixed points M and dS represent the
Minkowski and generalized de Sitter points as in the Bianchi I case
and retain the same stability character. The third point (CS) is the
curvature singularity\footnote{Strictly speaking, this points
represents at the same time the true past curvature singularity of
open Friedmann models (the left face of the sphere quarter in Fig.\
\ref{fig3}) as well as the coordinate singularity of the Milne
model, the line $X=Z=0$ in the same figure.}, which is curvature
dominated ($C \to 1$, or $c \to \infty$). The fourth point (IS)
represents the isotropic singularity and the final fixed point (KS)
represents the anisotropic Kasner singularity. We now discuss the
state space diagrams for this system. The case $\alpha>-1$ is shown
in Fig.~\ref{fig3}(a) (this corresponds to case {\bf A}). The $X-Z$
plane ($C=0$) is the invariant manifold which represents the reduced
and compactified Bianchi I state space, which is a manifold of the
larger Bianchi V state space. The $Z-C$ plane ($X=0$) is the vacuum
invariant manifold and the $X-C$ plane ($Z=0$) is the isotropic
invariant manifold. The Minkowski point M is the future attractor
for all trajectories in the region. The isotropic singularity (IS)
is the generic past attractor, while the anisotropic Kasner
singularity (KS) and curvature singularity (CS) have saddle
stability character. As in the Bianchi I case, in general the past
singularity is isotropic.

In the case $\alpha<-1$, depicted in Fig.~\ref{fig3}(b), there are
five fixed points, M, dS, CS, IS and KS.  The state space is split
into two regions by the separatrix manifold joining the points dS,
CS and KS. This manifold is given by $X=X_{\Lambda}$, where:
\begin{eqnarray}
X_{\Lambda} &=& \tau\sqrt{\frac{1-C^2-Z^2}{1+\tau^2}}.
\end{eqnarray}

\noindent and $\tau=-(\alpha+1)$. In the $Z-C$ plane ($X=0$, vacuum
invariant manifold) the dynamics are unaffected by the change in
$\alpha$. The $X-C$ plane ($Z=0$, isotropic invariant manifold) is
divided into two regions by the separatrix. In the region above the
separatrix ($X>X_{\Lambda}$, corresponding to case {\bf B}) the
fluid behaves in a standard manner. The IS fixed point has repeller
stability character and all trajectories are asymptotic to it in the
past. The dS fixed point is the generic future attractor. The CS and
KS fixed points have saddle stability character in this region. In
the region below the separatrix ($X<X_{\Lambda}$, corresponding to
case {\bf C}) the fluid behaves in a phantom manner. The KS fixed
point is a repeller  and the dS fixed point the attractor. The M and
CS fixed points have saddle stability character. In general,
trajectories are asymptotic to the Kasner singularity in the past
and the generalized de Sitter model in the future.

\section{Dynamics of anisotropy and inhomogeneity}\label{section4}

We have seen in the previous section that generic trajectories of
Bianchi I and V anisotropic models  with standard (non-phantom)
matter of type {\bf A} and {\bf B} asymptotically approach an
isotropic singularity in the past, with the Hubble normalized
shear vanishing, $\Sigma\rightarrow 0$. Since the Bianchi I and V
models are rather special and homogeneous, we now want to see if
this behavior is restricted to these models only, or if instead is
stable against small generic inhomogeneuos perturbations. That is,
we want to see if appropriately defined dimensionless quantities
measuring the deviation from homogeneity in a covariant gauge
invariant way decay to zero when the singularity is approached.
Since we are only interested in the neighborhood of the
singularity, we only need to consider perturbations of  the
spatially flat ($K=0$) homogenous and isotropic cosmological
models (the flat Friedmann models, i.e.\ the vertical $X$ axis in
Figs. \ref{fig2} and \ref{fig3}) with  quadratic EoS
\begin{equation}
P=\alpha\rho + \frac{\rho^2}{\rho_{c}}.
\end{equation}
Given this EoS, $\rho+3P>0$ is  always satisfied for high enough
$\rho$, whatever $\alpha$ is, hence at high enough energies the
background evolution is non-inflationary, $\ddot{a}<0$. Therefore
we only need to consider large scale perturbations, i.e.\ scales
much larger than the Hubble horizon $H^{-1}$, because for any
given perturbation scale $\lambda$ there is always a time $t_H$
(the horizon crossing time) such that for $t\ll t_H$ we have
$\lambda\gg H^{-1}$.

For simplicity we now focus on models of type {\bf A} only, with
$\alpha>-1$; it can be shown that similar results can be obtained
in case {\bf B}, $\alpha<-1$. For this case we re-write Eq.\
(\ref{rho1}) as
\begin{equation}
\rho(a)=\frac{\rho_{c}(\alpha+1)}{
(\frac{a}{a_s})^{3(\alpha+1)}-1},
\end{equation}
where
\begin{equation}
a_{s}\equiv A^{\frac{1}{3(\alpha+1)}}
\end{equation}
is the finite value of the scale factor such that the energy
density has a pole like singularity. $A$ is a constant given by
(\ref{AA}) and is expressed in terms of $\rho_{o}$, $a_{o}$,
$\rho_{c}$ and $\alpha$. Given that the background is a spatially
flat Friedmann model the Friedmann equation ($H^2=\rho/3$) gives
directly  the Hubble expansion scalar $H$, while $a_{o}$ can be
fixed arbitrarily, a freedom that we will use later. The sound
speed for the fluid is given by:
\begin{equation}
c_{s}^{2}=\frac{\partial P}{\partial \rho }=\alpha +
\frac{2\rho}{\rho_{c}}.
\end{equation}

\noindent The more commonly used EoS parameter $w$
is:
\begin{equation}
w=\frac{ P}{\rho }=\alpha + \frac{\rho}{\rho_{c}}.
\end{equation}

\noindent The sound speed and EoS variable are then
related by the following relationship:
\begin{equation}
c_{s}^{2}=2w -\alpha.
\end{equation}
Thus in case {\bf A} our parameter $\alpha$ could be replaced by
the combination $\alpha=2w-c_s^2$ of the more commonly used
parameters $w$ and $c_s^2$.

\subsection{Dimensionless variables and harmonics}

We use the covariant 1+3 approach ~\cite{EB,BDE,VEE} to study
inhomogeneous and anisotropic gauge-invariant perturbations. In
this approach, the information relating to inhomogeneous and
anisotropic perturbations is contained in a set of (1+3) covariant
variables. These variables are defined with respect to a preferred
time-like observer congruence $u_a$. We use appropriate
dimensionless density, expansion and curvature gradients which
describe the scalar and vector parts of the corresponding
perturbations:
\begin{eqnarray}
{\Delta}_{a} &=& \frac{a}{\rho}\nabla_{a}\rho, \\
Z_{a} &=& \frac{{\mathcal{Z}}_{a}}{H} = \frac{a}{H}\nabla_{a}H,\\
{\mathcal{C}}_{a} &=& {a^3}\nabla_{a} {}^{(3)}{R}.
\end{eqnarray}

\noindent The dimensionless expansion normalized quantities for
the shear $\sigma_{ab}$, the vorticity $\omega_a$, the electric
$E_{ab}$ and magnetic $H_{ab}$ parts of the Weyl tensor are:
\begin{equation}
\Sigma_{ab}=\frac{\sigma_{ab}}{H}\;,~~ W_a=\frac{\omega_a}{H}\;,~~
{\cal E}_{ab}=\frac{E_{ab}}{H^2}\;,~~ {\cal
H}_{ab}=\frac{H_{ab}}{H^2}\;,
\end{equation}

\noindent where $H$ is the Hubble scalar, $H=\dot{a}/a$. Here $\omega_a=\eta_{abc}\omega^{bc}$ is the
usual vorticity vector. In order to calculate all the relevant
tensor perturbations it is useful to define a set of variables
corresponding to the curls of the standard quantities. The curl of
a given quantity will be denoted by an over-bar.
\begin{equation}
\bar{\Sigma}_{ab}=\frac{1}{H} \textmd{curl}{\Sigma_{ab}}\;,~~
\bar{{\cal E}}_{ab}=\frac{1}{H}\textmd{curl}{\cal E}_{ab} \;,~~
\bar{\cal H}_{ab}=\frac{1}{H}\textmd{curl}{\cal H}_{ab}\;.
\end{equation}

\noindent The harmonics defined in~\cite{BDE} are used to expand
the above tensors $X_a\;,X_{ab}$ in terms of scalar (S), vector
(V) and tensor (T) harmonics $Q$. This results in a covariant and
gauge invariant splitting of the evolution and constraint
equations for the above quantities into three sets of evolution
and constraint equations for scalar, vector and tensor modes. The
scalar, vector and tensor quantities can be split such that:
\begin{eqnarray}
X&=&X^S Q^S \\
X_a&=&k^{-1}X^S Q_{a}^S+X^V Q_{a}^V \\
X_{ab}&=&k^{-2}X^S Q_{ab}^S+k^{-1}X^VQ_{ab}^V+X^T Q_{ab}^T\;.
\end{eqnarray}

\noindent As said above, we will carry out our analysis in the
long-wavelength limit $\lambda\gg H^{-1}$, or $k^2/a^2H^2\ll 1$
($\lambda=a/k$). This is analogous to dropping the Laplacian terms
in the evolution and constraint equations.

\begin{table*}[t]
\caption{\label{table1} The large scale behavior of the scalar
quantities. The constants of integration ($\Delta_{i}$)
correspond to the two independent modes. The non-zero constants
$B_{i}$, $Z_{i}$, $\mathcal{C}_{0}$, $\Sigma_{i}$ and ${\cal
E}_{i}$ can all be expressed in terms of $\Delta_{1,2}$,
$\rho_{c}$ and $\alpha$. The symbol $\Xi$ represents the generalized
hypergeometric function given in Eq.~(\ref{hyp}). The third column
gives the behavior of the perturbation at the singularity. The
last column gives the approximate late time behavior of the
perturbation. The limits are given under the assumption that
$\alpha>-1$ and $\rho_{c}>0$. }
\begin{ruledtabular}
\begin{tabular}{ccccc}
Quantity  & $X_{-}$ Mode & $X_{+}$ Mode & Limit($a \rightarrow a_{s}$) & Limit($a \gg a_{s}$)\\
\hline
 & & & & \\
$\Delta$       &  $\Delta_{1}
\frac{\sqrt{\rho_{c}(\alpha+1)a^{3(\alpha+1)}-{1}}}{a^3}$  &
$\Delta_{2} \frac{[{\rho_{c}(\alpha+1)a^{3(\alpha+1)}-1}]}{a^2}
\left( 1 + \frac{B_{1}}{a}\Xi \right) $ & $0$, $0$ & $a^{ {3(\alpha-1)}/{2}}$, $ a^{3\alpha+1} $ \\
& & & & \\
$Z$            & $Z_{1}
\frac{\sqrt{\rho_{c}(\alpha+1)a^{3(\alpha+1)}-{1}}}{a^3}$ & $Z_{2}
\frac{[{\rho_{c}(\alpha+1)a^{3(\alpha+1)}-1}]}{a^2} \left(
1 - \frac{B_{2}}{a}\Xi \right) $ & $0$, $0$ & $a^{ {3(\alpha-1)}/{2}}$, $ a^{3\alpha+1} $\\
& & & & \\
$\mathcal{C}$& 0 &  $\mathcal{C}_{0}$ & $0$, $\mathcal{C}_{0}$ & $0$, $\mathcal{C}_{0}$\\
& & & & \\
$\Sigma$       & $\Sigma_{1}
\frac{\sqrt{\rho_{c}(\alpha+1)a^{3(\alpha+1)}-{1}}}{a^3} $ &
$\Sigma_{2}\frac{[{\rho_{c}(\alpha+1)a^{3(\alpha+1)}-{1}}]}{a^2}
\left( 1 - \frac{B_{2}}{a}\Xi \right)$ & $0$, $0$ & $a^{ {3(\alpha-1)}/{2}}$, $ a^{3\alpha+1} $\\
& & & & \\
${\cal E} $    &  ${\cal E}_{1}
\frac{\sqrt{\rho_{c}(\alpha+1)a^{3(\alpha+1)}-{1}}}{a^3}$  &
${\cal E}_{2} \frac{[{\rho_{c}(\alpha+1)a^{3(\alpha+1)}-1}]}{a^2}
\left( 1 + \frac{B_{1}}{a}\Xi \right)$ & $0$, $0$ & $a^{ {3(\alpha-1)}/{2}}$, $ a^{3\alpha+1} $\\
& & & & \\
\end{tabular}
\end{ruledtabular}
\end{table*}

\subsection{Scalar perturbations}

The
evolution and constraint equations are greatly simplified if we
use the previously defined dimensionless time derivative:
\begin{eqnarray}
(~~)'=\frac{1}{H}\frac{d}{dt}=a\frac{d}{da}.
\end{eqnarray}

\noindent The linearized  evolution equations for scalar
perturbations are:
\begin{eqnarray}
{\Delta}' &=& 3w\Delta - \frac{(w+1)}{H}\mathcal{Z}, \\
{\mathcal{Z}}' &=& -2\mathcal{Z} - \frac{3H}{2}\Delta,\\
E' &=& -3E -\frac{3H}{2}(w+1)\sigma, \\
\sigma' &=& -2\sigma -\frac{E}{H}.
\end{eqnarray}

\noindent The set of equations contain a mixture of Hubble
normalized and non-normalized variables, however all solutions
presented are appropriately normalized. In a addition we have a
set of constraint equations for the scalar quantities:
\begin{eqnarray}
\mathcal{C} &=& -4 a^{2} H \mathcal{Z} + 2 a^{2} \rho \Delta,\\
3\Sigma &=& 2Z, \\
{\cal E} &=&\Delta.
\end{eqnarray}

\noindent The system of equations is particularly difficult to
solve in the above form. This is due to the form of $\rho(a)$,
$H(a)$ and $w(a)$ (the pole like behavior at $a=a_{s}$). In order
to solve the system of equations we carry out a change of
variables. This is a parameter specific
($\rho_{c}$ and $\alpha$) non-linear transformation of the scale
factor. Once the solutions are obtained, they are again expressed
in terms of the original normalized scale factor $a$. The
solutions to the system of equations can be split into modes which
are sourced by the gradient of the 3-curvature ($\mathcal{C}$,
which is constant in the long-wavelength limit) perturbation
(subscript $+$) and those which are not (subscript $-$). The
solutions for the scalar contribution to the density and expansion
gradient's are of the form:
\begin{eqnarray}
\Delta_{-} &=& \Delta_{1}\frac{\sqrt{\rho_{c}(\alpha+1)a^{3(\alpha+1)}-{1}}}{a^3}  ,  \\
\nonumber\\
\Delta_{+} &=& \Delta_{2}\frac{[{\rho_{c}(\alpha+1)
a^{3(\alpha+1)}-{1}}]}{a^2} \times \left( 1 +  \frac{B_{1}}{a}\Xi \right) ,\\
\nonumber\\
{Z}_{-} &=& {Z}_{1}\frac{\sqrt{\rho_{c}(\alpha+1)a^{3(\alpha+1)}-{1}}}{a^3}  ,  \\
\nonumber\\
{Z}_{+} &=& {Z}_{2}\frac{[{\rho_{c}(\alpha+1)
a^{3(\alpha+1)}-{1}}]}{a^2} \times \left( 1 -  \frac{B_{2}}{a}\Xi
\right) .
\end{eqnarray}

\noindent The $\Delta_{i}$'s ($i=1,2$) are the arbitrary constants of
integration and correspond to the two independent modes. The other
constants $B_{i}$ and $Z_{i}$ can be expressed in terms of the
arbitrary constants $\Delta_{i}$ and the EoS parameters
$\rho_{c}$ and $\alpha$. The symbol $\Xi$ represents the generalized
hypergeometric function and is of the form:
\begin{eqnarray}\label{hyp}
\Xi &=& {}_{2}{\mathcal{F}}_{1}\left(
\left[\frac{1}{2},\frac{3\alpha+2}{3(\alpha+1)} \right],
\left[\frac{3}{2}\right],1-b\right),\\
\nonumber\\
b &=&\rho_{c}(\alpha+1)a^{3(\alpha+1)}.
\end{eqnarray}

\noindent The generalized hypergeometric function is equal to
unity at the singularity (${}_{2}{\mathcal{F}}_{1}([i,j],[k],0)=1
$), where also $b=1$ after using the freedom in $a_o$ to set
$A=[\rho_c(1+\alpha)]^{-1}$.  In addition the quantity
$(1\pm~B_{i}\Xi/a) \to 1$ as $a\gg a_{s}$, the hypergeometric
function is a correction to the solutions at the singularity and
plays a subdominant role at late times. The perturbations all
decay as we approach the singularity in the past ($\Delta_{\pm}\to
0$ and $Z_{\pm} \to 0$ as $a \to a_{s}$). At late times ($a\gg
a_{s}$) the various modes are approximately:
\begin{eqnarray}
\Delta_{-} &\approx& \Delta_{1} a^{ {3(\alpha-1)}/{2}}, \\
\nonumber\\
\Delta_{+} &\approx& \Delta_{2} a^{3\alpha+1} ,\\
\nonumber\\
{Z}_{-} &\approx& {Z}_{1} a^{ {3(\alpha-1)}/{2}} ,  \\
\nonumber\\
{Z}_{+} &\approx& {Z}_{2} a^{3\alpha+1} .
\end{eqnarray}

\noindent This is the corresponding behavior one would expect for
a fluid with a linear EoS. The modes asymptotically evolve as
there linear EoS counterparts at late times. The perturbation
modes for each of the scalar quantities can be found in
Table~\ref{table1}. The  constants of integration are the
$\Delta_{i}$'s and all other constants ($Z_{i}$, $\Sigma_{i}$,
$\mathcal{E}_{i}$, $B_{i}$ and $\mathcal{C}_{0}$ ) can be
expressed in terms of these constants and the EoS parameters. The
behavior of each mode is also given in the two limits, the first
is at the singularity ($a \to a_{s}$) and the second is at late
times ($a\gg a_{s}$). The limits are given assuming $\alpha>-1$
and $\rho_{c}>0$.

\begin{table*}[t]
\caption{\label{table2} The large scale behavior of the vector
quantities. The  constants of integration ($\Delta_{i}$ and
$W_{1}$) correspond to the three independent modes. The non-zero
constants $B_{i}$, $Z_{i}$, $\mathcal{C}_{0}$, $\mathcal{H}_{1}$,
$\Sigma_{i}$ and ${\cal E}_{i}$ can all be expressed in terms of
$\Delta_{1,2}$, $W_{1}$, $\rho_{c}$ and $\alpha$. See the caption
of Table \ref{table1} for further details. }
\begin{ruledtabular}
\begin{tabular}{ccccc}
Quantity  & $X_{-}$ Mode & $X_{+}$ Mode & Limit($a \rightarrow a_{s}$) & Limit($a \gg a_{s}$)\\
\hline
 & & & & \\
$\Delta$       &  $\Delta_{1}
\frac{\sqrt{\rho_{c}(\alpha+1)a^{3(\alpha+1)}-{1}}}{a^3}$  &
$\Delta_{2} \frac{[{\rho_{c}(\alpha+1)a^{3(\alpha+1)}-1}]}{a^2}
\left( 1 + \frac{B_{1}}{a}\Xi \right) $ & $0$, $0$ & $a^{ {3(\alpha-1)}/{2}}$, $ a^{3\alpha+1} $ \\
& & & & \\
$Z$            & $Z_{1}
\frac{\sqrt{\rho_{c}(\alpha+1)a^{3(\alpha+1)}-{1}}}{a^3}$ & $Z_{2}
\frac{[{\rho_{c}(\alpha+1)a^{3(\alpha+1)}-1}]}{a^2} \left(
1 - \frac{B_{2}}{a}\Xi \right) $ & $0$, $0$ & $a^{ {3(\alpha-1)}/{2}}$, $ a^{3\alpha+1} $\\
 & & & & \\
$\mathcal{C}$& 0 &  $\mathcal{C}_{0}$ & $0$, $\mathcal{C}_{0}$ & $0$, $\mathcal{C}_{0}$\\
 & & & & \\
$W$       & $W_{1} \frac{
[\rho_{c}(\alpha+1)a^{3(\alpha+1)}-{1}]^{{5}/{2}} }{a^{3\alpha+8}}
$
& $-$  &  $0$&  $a^{ {(9\alpha-1)}/{2}}$ \\
 & & & & \\
$\mathcal{H}$       & $\mathcal{H}_{1} \frac{
[{\rho_{c}(\alpha+1)a^{3(\alpha+1)}-{1}}] }{a^{4}} $
& $-$  & $0$ &  $a^{ {3(\alpha-1)}}$ \\
 & & & & \\
$\Sigma$       & $\Sigma_{1}
\frac{\sqrt{\rho_{c}(\alpha+1)a^{3(\alpha+1)}-{1}}}{a^3} $ &
$\Sigma_{2}\frac{[{\rho_{c}(\alpha+1)a^{3(\alpha+1)}-{1}}]}{a^2}
\left( 1 - \frac{B_{2}}{a}\Xi \right)$ & $0$, $0$ & $a^{ {3(\alpha-1)}/{2}}$, $ a^{3\alpha+1} $\\
& & & & \\
${\cal E} $    &  ${\cal E}_{1}
\frac{\sqrt{\rho_{c}(\alpha+1)a^{3(\alpha+1)}-{1}}}{a^3}$  &
${\cal E}_{2} \frac{[{\rho_{c}(\alpha+1)a^{3(\alpha+1)}-1}]}{a^2}
\left( 1 + \frac{B_{1}}{a}\Xi \right)$ & $0$, $0$ & $a^{ {3(\alpha-1)}/{2}}$, $ a^{3\alpha+1} $\\
 & & & & \\
\end{tabular}
\end{ruledtabular}
\end{table*}

\subsection{Vector perturbations}

The linearized evolution equations for the vector components of
the perturbations in the large-scale limit are:
\begin{eqnarray}
{\Delta}' &=& 3w\Delta - \frac{(w+1)}{H}\mathcal{Z}, \\
{\mathcal{Z}}' &=& -2\mathcal{Z} - \frac{3H}{2}\Delta, \\
\omega' &=& (3c_{s}^{2} - 2)\omega ,\\
E' &=& -3E -\frac{3H}{2}(w+1)\sigma, \\
\sigma' &=& -2\sigma -\frac{E}{H}.
\end{eqnarray}

\noindent The constraint equations for the vector variables are
\begin{eqnarray}
\mathcal{C}&=&-4 a^{2}H\mathcal{Z}+2a^{2}\rho\Delta ,\\
{\cal H} &=& \frac{3a}{2} (w+1) H W,\\
3\Sigma &=& 2Z, \\
{\cal E} &=&\Delta.
\end{eqnarray}

\noindent The perturbation modes for each of the vector quantities
can be found in Table~\ref{table2}. The  constants of
integration ($\Delta_{i}$ and $W_{1}$) correspond to the three
independent modes. The non-zero constants $B_{i}$, $Z_{i}$,
$\mathcal{C}_{0}$, $\mathcal{H}_{1}$, $\Sigma_{i}$ and ${\cal
E}_{i}$ can all be expressed in terms of the constants of
integration and the EoS parameters. The last two columns give the
behavior of the perturbation in the two limits. As in the case of
the scalar contributions all perturbations decay at the
singularity and asymptotically approach the standard linear EoS
behavior at late times.

\begin{table*}[t]
\caption{\label{table3} The large scale behavior of the tensor
quantities. The  constants of integration ($\Sigma_{i}$ and
$\mathcal{H}_{i}$) correspond to the four independent modes. The
non-zero constants $B_{i}$, $\mathcal{E}_{i}$, $\bar{{\cal
E}}_{i}$ and $\bar{\Sigma}_{i}$ can all be expressed in terms of
$\Sigma_{1,2}$, $\mathcal{H}_{1,2}$, $\rho_{c}$ and $\alpha$. See
the caption of Table \ref{table1} for further details.  }
\begin{ruledtabular}
\begin{tabular}{ccccc}
Quantity  & $X_{-}$ Mode & $X_{+}$ Mode & Limit($a \rightarrow a_{s}$) & Limit($a \gg a_{s}$)\\
\hline
 & & & & \\
$\Sigma$       & $\Sigma_{1}
\frac{\sqrt{\rho_{c}(\alpha+1)a^{3(\alpha+1)}-{1}}}{a^3} $ &
$\Sigma_{2}\frac{[{\rho_{c}(\alpha+1)a^{3(\alpha+1)}-{1}}]}{a^2}
\left( 1 - \frac{B_{2}}{a}\Xi \right)$ & $0$, $0$ & $a^{ {3(\alpha-1)}/{2}}$, $ a^{3\alpha+1} $\\
& & & & \\
${\cal E} $    &  ${\cal E}_{1}
\frac{\sqrt{\rho_{c}(\alpha+1)a^{3(\alpha+1)}-{1}}}{a^3}$  &
${\cal E}_{2} \frac{[{\rho_{c}(\alpha+1)a^{3(\alpha+1)}-1}]}{a^2}
\left( 1 + \frac{B_{1}}{a}\Xi \right)$ & $0$, $0$ & $a^{ {3(\alpha-1)}/{2}}$, $ a^{3\alpha+1} $\\
 & & & & \\
${\cal H} $    & ${\cal H}_{1}
\frac{[{\rho_{c}(\alpha+1)a^{3(\alpha+1)}-1}]}{a^4}$ & ${\cal
H}_{2}
\frac{{[{\rho_{c}(\alpha+1)a^{3(\alpha+1)}-1}]}^{\frac{3}{2}}}{a^3}
\left( 1 - \frac{B_{3}}{a}\Xi \right)$ & $0$, $0$ & $a^{ {3\alpha-1}}$, $ a^{3(3\alpha+1)/2} $\\
 & & & & \\
$\bar{{\cal E}} $    & $\bar{{\cal E}}_{1}
\frac{[{\rho_{c}(\alpha+1)a^{3(\alpha+1)}-1}]}{a^4}$ & $\bar{{\cal
E}}_{2}
\frac{{[{\rho_{c}(\alpha+1)a^{3(\alpha+1)}-1}]}^{\frac{3}{2}}}{a^3}
\left( 1 + \frac{B_{4}}{a}\Xi \right)$ & $0$, $0$ & $a^{ {3\alpha-1}}$, $ a^{3(3\alpha+1)/2} $\\
 & & & & \\
$\bar{\Sigma} $    & $\bar{\Sigma}_{1}
\frac{[{\rho_{c}(\alpha+1)a^{3(\alpha+1)}-1}]}{a^4}$ &
$\bar{\Sigma}_{2}
\frac{{[{\rho_{c}(\alpha+1)a^{3(\alpha+1)}-1}]}^{\frac{3}{2}}}{a^3}
\left( 1 - \frac{B_{3}}{a}\Xi \right)$ & $0$, $0$ & $a^{ {3\alpha-1}}$, $ a^{3(3\alpha+1)/2} $\\
 & & & & \\
$\bar{{\cal H}} $    & $0$ & $0$ & $-$ & $-$\\
 & & & & \\
\end{tabular}
\end{ruledtabular}
\end{table*}

\subsection{Tensor perturbations}

The linearized evolution equations for the tensor components of
the perturbations in the large-scale limit are:
\begin{eqnarray}
E' &=& -3E -\frac{3H}{2}(w+1)\sigma, \\
\sigma' &=& -2\sigma -\frac{E}{H}, \\
{\cal H}' &=& -3{\cal H} - \frac{1}{H} \bar{E}  \\
\bar{E}' &=& -3\bar{E} -\frac{3H}{2}(w+1){\cal H}.
\end{eqnarray}

\noindent The constraint equations for the tensor variables are
\begin{eqnarray}
\cal{H}&=& \bar{\Sigma} ,\\
\bar{{\cal H}} &=& 0.
\end{eqnarray}

\noindent The perturbation modes for each of the tensor quantities
can be found in Table~\ref{table3}. The  constants of
integration ($\Sigma_{i}$ and $\mathcal{H}_{i}$) correspond to the
four independent modes. The non-zero constants $B_{i}$,
$\mathcal{E}_{i}$, $\bar{{\cal E}}_{i}$ and $\bar{\Sigma}_{i}$ can
all be expressed in terms of the constants of integration and the
EoS parameters. The perturbations all decay at the singularity and
evolve as there linear EoS counterparts at late times.

\section{Conclusions}

In this paper we have investigated the effects of a quadratic EoS
in homogenous and inhomogeneous anisotropic cosmological models in
general relativity, in order to understand if in this context the
quadratic EoS can isotropize the universe at early times, when the
initial singularity is approached.

The first motivation for this work is that in the context of brane
world cosmological models the  term quadratic in the energy
density that appears in the effective 4-dimensional equations of
motion on the brane dominates at early times, driving the
evolution to an initial isotropic singularity under reasonable
assumptions. Therefore, under these assumptions,  in the brane
world scenario we do not have the classical isotropy problem: the
isotropy we observe today is built in, that is it follows from
{\it generic initial conditions}. This is at odds with the
classical and now widely accepted results in  general relativity
\cite{BKL,LL,Ashtekar}, where general cosmological models with a
linear EoS are highly anisotropic in the past, approaching the
singularity {\it a-la} BKL, i.e. with mixmaster dynamics, so that
in general one needs {\it special initial conditions} to have
isotropy at late times, even to initiate an inflationary phase
\cite{KT}.

Secondly, we also wanted to study the effects at early times and
at high energies of the same type of EoS used in Paper
I\,\cite{AB}, relaxing the symmetries assumed there. The focus in
Paper I\,\cite{AB} was on the possible use of a quadratic EoS as
an effective way of representing a dark energy component or
unified dark matter in the context of homogeneous isotropic models
of relevance for the universe at late times. Here we have used the
simplified EoS $P=\alpha \rho + \rho^2/\rho_c$, which still allows
for an effective cosmological constant and phantom behavior, and
is general enough to analyze the dynamics at high energies.

In summary, our aim here was to investigate if  the behavior of
the anisotropy at the singularity found in the brane scenario is
recreated in the general relativistic context if we consider a
quadratic term in the EoS. That is to say, in the language of
dynamical system theory, do we have a scenario where  the generic
asymptotic past attractor in general relativity is an isotropic
model, thanks to a non-linear (specifically, quadratic) EoS.

To this end, we have first summarized  in Section \ref{section2}
the classification of the energy density evolution in three
classes, {\bf A}, {\bf B} and {\bf C}, following from the energy
conservation for the EoS $P=\alpha \rho +  \rho^2/\rho_c$  (a
broader classification  can be found in Section II of Paper
I\,\cite{AB}). In Section \ref{section3}, using dynamical system
methods \cite{WE,AP},  we have studied the Bianchi I and V
cosmological models with a fluid with this EoS. We first
considered the case when $\alpha>-1$ (case {\bf A}). The
physically interesting quantity $\Sigma$, the Hubble normalized
dimensionless  measure of shear, describes the anisotropy. In the
expanding Bianchi I and V models this quantity always decreases at
late times, as in the case with a {\it linear} EoS, approaching a
Minkowski model. However, we can now have stages at early times in
which $\Sigma$ grows,  vanishing at the initial singularity, and
generic trajectories in the compactified state space
asymptotically approach in the past the isotropic fixed point IS,
as depicted in Figs. \ref{fig2} and \ref{fig3}. This behavior is
similar to the quasi-isotropic expansion seen in the anisotropic
brane-world models, but occurs for different reasons. The
quasi-isotropic expansion is caused by the fact that the EoS
becomes ``super-stiff" at early times but decreases to finite
values at late times. At the initial singularity (which is of Type
III, see \cite{NOT, barrow}) the anisotropy is always sub-dominant
and we recover the standard {\it linear} EoS behavior at late
times. In the case $\alpha<-1$, the state space can be divided in
two regions. In the standard fluid region ($X>X_{\Lambda}$, case
{\bf B}) the behavior at early times is similar to the $\alpha>-1$
case, with generic trajectories in the compactified state space
approaching the isotropic fixed point IS of Figs. \ref{fig2} and
\ref{fig3}. At late times  instead trajectories  asymptotically
approach generalized expanding de Sitter models. Finally, in the
phantom fluid region ($X<X_{\Lambda}$, case {\bf C}) the
anisotropy dominates at early times and the past singularity is
anisotropic. Trajectories are asymptotic to the expanding de
Sitter model at late times.

The submanifold $\Sigma=0$ in the state space for  the Bianchi
models represents homogeneous isotropic spatially flat
Friedmann-Robertson-Walker models, with IS as past attractor. In
order to see if IS can be regarded as the past attractor of a
larger class of general  models with no symmetries,  we have
studied in Section \ref{section4}, using covariant and gauge
invariant variables \cite{EB,BDE,VEE},  the first order
inhomogeneous anisotropic perturbations of these flat isotropic
models, focusing on those of class {\bf A}. The linearized
equations have been solved in the large scale limit $\lambda\gg
H^{-1}$, neglecting the Laplacian operators terms. As explained in
Section \ref{intro}, only perturbations larger then the Hubble
horizon $H^{-1}$ are relevant in our analysis. The main results of
this section are given in Table's~\ref{table1},~\ref{table2} and
~\ref{table3}. The tables contain the form of the solutions and
asymptotic behavior of the physically relevant quantities for both
early ($a \to a_{s}$) and late times ($a \gg a_{s}$). The
quantities have been harmonically decomposed into the standard
scalar, vector and tensor components. These include the expansion
normalized vorticity, shear, and electric and magnetic
contributions of the Weyl tensor. In addition, we consider the
gradients of the energy density, expansion and three-curvature. As
expected, at late times the perturbations evolve as if the fluid
obeyed a {\it linear} EoS, i.e. the standard behavior is recovered
in this limit. The evolution of the perturbations at early times
is our main result. Indeed, we find that all perturbations tend to
zero near the singularity. Thus a perturbed quasi-isotropic model
with generic initial condition evolves from the isotropic past
attractor IS. This indicates that in general relativity with
matter described by a non-linear EoS  there are general
cosmological models, corresponding to a non zero measure subset of
all possible generic initial conditions \cite{LL}, that isotropize
as the initial singularity is approached, with IS as the past
attractor. As for the brane models, this would be at odds with the
standard BKL results  \cite{BKL,LL,Ashtekar}.

Singularities are commonly regarded as signaling the failure of
classical gravitational theories, requiring a quantum theory of
gravity to deal with trans-Plankian energies. Even in effective
theories, perhaps cosmology doesn't require them, as in the
pre-big-bang scenario \cite{pbb}. In any case, it is important to
understand which is the typical classical dynamical behavior in
the neighborhood of singularities,  as this possibly signals what
the generic outcome of a quantum theory should be at low
(sub-Plankian) energies. In the BKL picture for fluids with linear
EoS and other simple type of matter \cite{Ashtekar} this behavior
is mixmaster and the effects of matter become negligible close to
the singularity: we may say that effectively in this limit  the
space-time still tells matter how to move, but matter is no longer
able to tell the space-time how to curve. Perhaps this behavior
should be regarded as pathological, and peculiar of the
energy-momentum tensor of linear EoS matter (or other simple
energy-momentum tensor) in general relativity. In practice we know
close to nothing about matter fields at the highest (close to
Plankian) cosmological  energies, and these may require a
non-linear effective EoS description in general, or maybe Einstein
equations with an effective non-linear EoS are the Einstein frame
version of some effective low energy (sub-Plankian)  theory. In
both cases, our analysis  of Einstein equations with a fluid with
a quadratic EoS indicates that the typical initial state is
isotropic. The singularity is matter dominated, and of a type that
could admit an extension. Interestingly, the EoS considered  here
are a subclass of those used  in Paper I\,\cite{AB} to model a
dark energy component or unified dark matter, allowing for
effective cosmological constants.

\acknowledgments KNA is supported by PPARC (UK). MB is partly
supported by a visiting grant by MIUR (Italy). The authors would
like to thank Chris Clarkson, Roy Maartens and David Wands for
useful comments and discussions.


\end{document}